\documentclass[aps,pre,twocolumn,amsmath,amssymb,amsfonts,floatfix,superscriptaddress,showkeys,showpacs]{revtex4-1}
\usepackage{dcolumn}  
\usepackage{multirow}
\usepackage[T1]{fontenc}
\usepackage[latin1]{inputenc}
\usepackage{verbatim}
\usepackage{bm} 
\usepackage{hyperref}
\usepackage{natbib}
\usepackage[english]{babel}
\usepackage{babelbib}
\usepackage{graphicx}
\usepackage{epstopdf}
\usepackage{ifpdf} 
\usepackage{epsfig}
\usepackage{color}
\usepackage[usenames,dvipsnames]{xcolor}
\usepackage{mathrsfs}

\newcommand{\rme}{{\mathrm{e}}}

\newcommand{\dist}{|\mathbf{r}-\mathbf{r}'|}

\begin{document}

\title{Role of metallic core for the stability of virus-like particles in strongly coupled electrostatics}

\author{Leili \surname{Javidpour}}
\affiliation{School of Physics, Institute for Research in Fundamental Sciences (IPM), Tehran 19395-5531, Iran}
\author{An\v{z}e \surname{Lo\v{s}dorfer Bo\v{z}i\v{c}}}
\affiliation{Department of Theoretical Physics, Jo\v zef Stefan Institute, SI-1000 Ljubljana, Slovenia}
\author{Rudolf \surname{Podgornik}}
\affiliation{School of Physical Sciences and Kavli Institute for Theoretical Sciences, University of Chinese Academy of Sciences, Beijing 100049, China}
\affiliation{CAS Key Laboratory of Soft Matter Physics, Institute of Physics, Chinese Academy of Sciences (CAS), Beijing 100190, China}
\affiliation{Department of Physics, Faculty of Mathematics and Physics, University of Ljubljana, 1000 Ljubljana, Slovenia}
\affiliation{Department of Theoretical Physics, Jo\v zef Stefan Institute, SI-1000 Ljubljana, Slovenia}
\author{Ali \surname{Naji}}
\affiliation{School of Physics, Institute for Research in Fundamental Sciences (IPM), Tehran 19395-5531, Iran}

\begin{abstract}
We investigate the osmotic (electrostatic) pressure acting on the proteinaceous shell of a generic model of virus-like particles (VLPs), comprising a charged outer shell and a metallic nanoparticle core, coated by a charged layer and bathed in an aqueous electrolyte (salt) solution. Motivated by the recent studies accentuating the role of multivalent ions for the stability of VLPs, we focus on the effects of multivalent cations and anions in an otherwise monovalent ionic bathing solution. We perform extensive Monte-Carlo simulations based on appropriate Coulombic interactions that consistently take into account the effects of salt screening, the dielectric polarization of the metallic core, and the strong-coupling electrostatics due to the presence of multivalent ions. We specifically study the intricate roles these factors play in the electrostatic stability of the model VLPs. It is shown that while the insertion of a metallic nanoparticle by itself can produce negative, inward-directed, pressure on the outer shell, addition of only a small amount of multivalent counterions can robustly engender negative pressures, enhancing the VLP stability across a wide range of values for the system parameters.
\end{abstract}

\maketitle

\section{Introduction}

Encapsidation of non-biological cargo in viral proteinaceous capsids has attracted a lot of interest in recent years, connected with their role as (noninfectious) virus-like particles (VLPs) in applications such as gene transfer, drug delivery \cite{DragneaRev}, engineering of modern vaccine platforms  \cite{Manchester2010,Chackerian2016}, as well as in biomedical imaging and therapeutics applications~\cite{Pokorski2011,Steinmetz2010,Steinmetz2011,ReviewDing2018}.
Encapsidation of metallic nanoparticles such as gold and iron-oxide cores \cite{Dragnea2012} in viral capsids has  extensively been studied using experimental methodology \cite{DragneaPNAS2007,Dragnea2012,Loo2006,Loo2007,Mieloch2018}, and (to a lesser extent) using theoretical and computer simulation models \cite{HaganTheo2009,vdSchoot2015}. Nanoparticle encapsidation is typically done through origin-of-assembly templating or polymer templating  \cite{Pokorski2011}. In the first scenario, the nanoparticle core is decorated by an origin-of-assembly site that initiates the binding of the coat proteins and drives the self-assembly around the core. In the second scenario, the nanoparticle core is decorated with negatively charged polymers intended to mimic the effects due to the negative charge of the native viral cargo, which is the (highly negatively charged) nucleic-acid (DNA or RNA) genome.

The interactions between the viral capsid and its native genomic cargo can include both non-specific and highly specific  interactions, depending on the type of the virus. This consequently influences whether the encapsidation of artificial cargo  can be achieved without help from nucleic acids, or whether oligonucleotides of a given length and specific sequence are  needed for proper assembly~\cite{DragneaRev,Loo2006,Loo2007,DragneaACSNano2010,DragneaNanoLett2006,Aniagyei2009}.
A prime example of a virus that can serve in a variety of  functional nanoparticle assemblies  is the brome mosaic virus (BMV)~\cite{Yildiz2012}, while other ssRNA plant viruses, 
such as red clover necrotic mottle virus (RCNMV)  \cite{Loo2006,Loo2007} and cowpea chlorotic mottle virus (CCMV) \cite{Aniagyei2009} have also been used as nanoparticle containers.
The capsids of these viruses usually carry hypotopal protein N-terminal tails with a high positive charge, which, in many cases, bind with the genome at least partially through non-specific electrostatic interactions~\cite{DragneaRev}. Thus, nanoparticles decorated with negatively charged polymers may in principle mimic the electrostatic behavior of the genome and initiate the self-assembly of the viral capsid around the artificial core.

The chemical and physical properties of the non-biological VLP core play a major role in determining the efficiency of the formation of the proteinaceous shell around it, as well as in determining the VLP's overall stability and electrostatic properties~\cite{DragneaPNAS2007}. Recent experiments were successful in decoupling the role of charge and size of the encapsulated cargo in the assembly of VLPs~\cite{DragneaACSNano2010}. It was observed that there is a critical charge density of the core below which the VLPs do not form, even if it can otherwise fit well into the cavity of the proteinaceous shell and the total charge on the core is sufficient to completely neutralize the positively charged N-tails of a complete viral capsid.

The role of the bathing ionic solution conditions for the electrostatic stability of viruses and VLPs has also been studied in recent years \cite{HaganTheo2009}, primarily based on the mean-field Poisson-Boltzmann (PB) framework \cite{Israelachvili,VO}. It is however known from more recent developments \cite{holm,hoda_review,Naji_PhysicaA,Shklovs02,Levin02,perspective} that even a small amount of multivalent ions in the system can dramatically shift the governing electrostatic paradigm, from that described by the PB theory to a conceptually different paradigm known as the strong coupling (or, in its generalized form in the case of an ionic mixture, the dressed multivalent-ion) theory \cite{perspective}. This latter situation is believed to occur in many biologically relevant examples \cite{holm,book,PhysToday}. The counter-intuitive like-charge attraction is a major manifestation of strong-coupling interactions mediated by multivalent ions between macromolecular surfaces \cite{holm,hoda_review,Naji_PhysicaA,Shklovs02,Levin02,perspective}, and it  is considered to underlie exotic phenomena such as formation of large DNA condensates \cite{Bloom2,Yoshikawa1,Yoshikawa2,Pelta} and large bundles of microtubules \cite{Needleman} and F-actin \cite{Angelini03,Tang}. Multivalent ions are also known to play a key role in dense DNA packaging in viruses and nano-capsids \cite{Plum,Raspaud,Savithri1987,deFrutos2005,Siber,Evilevitch2008,Evilevitch2006,Evilevitch2003}
although their effects are still not completely understood.

Motivated by these developments and the relevance of electrostatic interactions for the stability of the VLP formation, we investigate the effects of multivalent cations and anions in an otherwise monovalent ionic bathing solution, in which a model VLP is formed. We formulate a theoretical approach that can take into account also the effects of the metallic nanoparticle core polarization due to the discontinuous jump of the dielectric constant at its surface in the presence of ionic screening, as well as the strong-coupling  electrostatic effects generated by the presence of solution multivalent ions. In particular we address also the combined effect of dielectric images, due to the sharp increase of the dielectric constant at the metallic core surface, as well as the dielectric images which result in the inhomogeneous distribution of the salt that cannot penetrate inside the metallic core  (see also the recent work in Ref. \cite{Matej2018}). The description of electrostatic interactions between the solution components and the metallic core, as well as the electrostatic interactions between the ionic solution components in the presence of the core is then used in Monte-Carlo simulations to compute the equilibrium osmotic pressure exerted on the proteinaceous shell in presence and absence of the metallic nanoparticle core.  We specifically address  the change of sign in this pressure as a result of the strong coupling electrostatics in the presence of sharp dielectric boundaries for model VLPs in the presence and absence of the metallic core.

\section{Model and Methods}

\subsection{Model: Geometry and basic features}

\begin{figure}[t!]
\includegraphics[width=4.5cm]{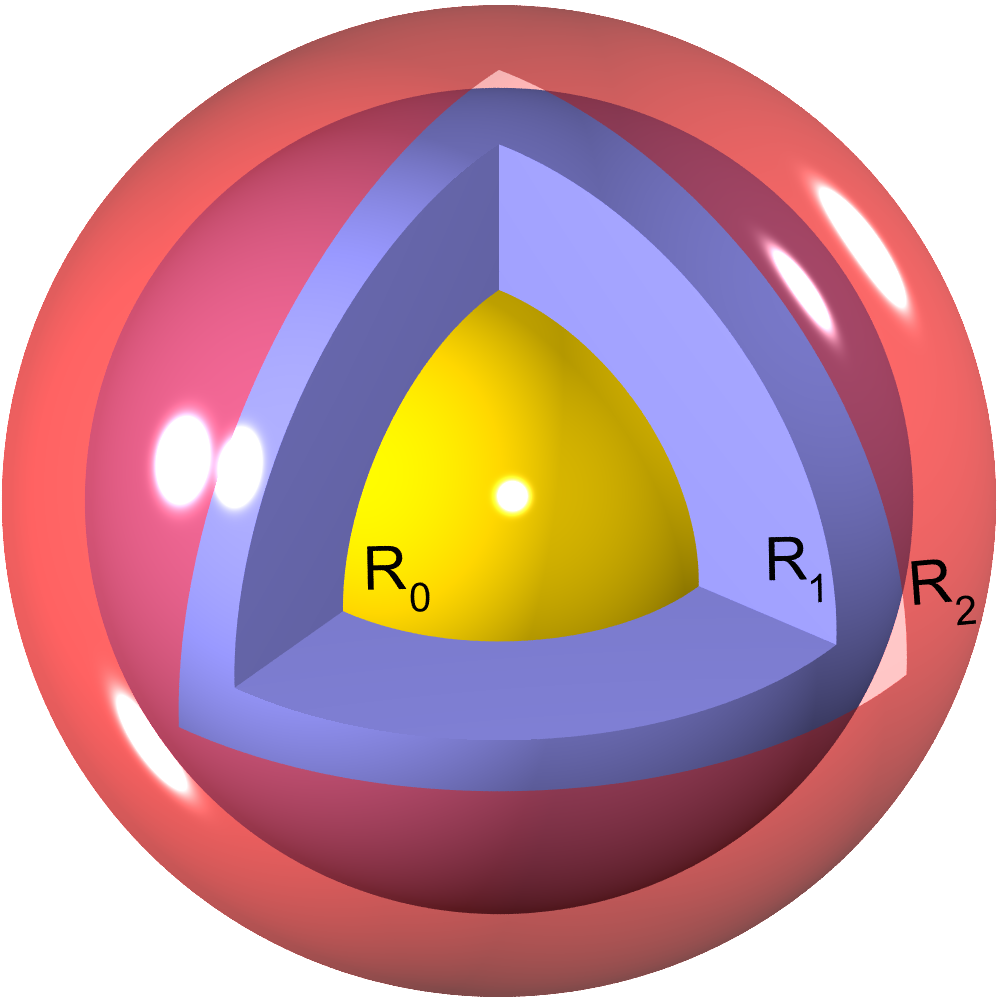}
\caption{Schematic view of a model VLP comprising an electroneutral metallic (yellow) core of radius $R_0$, coated by a (blue) layer of outer radius $R_1$ and surface charge density $\sigma_1$, encapsidated in a charged (red) shell of radius $R_2$ and surface charge density $\sigma_2$ (see the text for details). For the sake of illustration, parts of the outer shell and the coating layer are removed in the schematic picture to show the core region.}
\label{fig:sketch}
\end{figure}

We model the metallic nanoparticle (NP) core as an ideally polarizable and electrostatically neutral sphere of radius $R_0$, decorated with a coating layer of outer radius $R_1$ (representing, e.g., a polymeric layer such as the negatively charged polyethylene glycol used to cover encapsidated gold NPs~\cite{DragneaPNAS2007,DragneaACSNano2010,DragneaNanoLett2006}). The coating layer in general bears a surface charge density of $\sigma_1e_0$, where $e_0$ is the elementary charge.
The enclosing capsid will likewise be treated as a spherical shell of radius $R_2$ and surface charge density $\sigma_2e_0$ (see Fig.~\ref{fig:sketch}).
For simplicity, we shall refer to the charge distribution of the coating layer  and that of the capsid as the ``outer'' and ``inner'' charged shells.
While the metallic core is assumed to be strictly impermeable, the capsid and the coating layer are assumed to be permeable to water molecules and monovalent solution ions, which can thus be present within the whole region $r>R_0$. The (generally larger) multivalent ions are assumed to permeate inside the shell but not within the coating $r<R_1$. As such, we take the region outside the metallic core as a medium of uniform dielectric constant. The bathing solution is assumed to contain a base monovalent salt (1:1) and an additional asymmetric ($q:1$) salt of bulk concentrations $n_0$ and $c_0$, respectively.

When multivalent ions have a relatively large valency $q$ (typically $|q|> 2$, as exemplified by polyamines \cite{PolyaminesViruses,PolyaminesSize}, e.g., tri- and tetravalent spermidine and spermine, and ionic complexes, e.g., trivalent cobalt hexammine \cite{GelbartPRL2011}), the ionic mixture can no longer be treated using traditional mean-field approaches (applicable to monovalent ions)  \cite{holm}, nor using standard strong-coupling methods (applicable to multivalent counterion-only solutions) \cite{hoda_review,Naji_PhysicaA,Shklovs02,Levin02}, as one must concurrently account for the weak- and strong-coupling nature of interactions mediated by mono- and multivalent ions, respectively \cite{perspective}. The dressed multivalent-ion approach~\cite{SCdressed1} provides one such framework in the case of highly asymmetric ionic mixtures and in a relatively broad range of salt concentrations (with $c_0$ typically being less than few tens of mM). Having been successfully tested against simulations \cite{SCdressed2,SCdressed3,leili1,leili2} and recent experiments \cite{Trefalt} within its predicted regime of validity, it can offer considerable  nontrivial  simplification in the study of highly asymmetric ionic mixtures by systematically integrating out the  monovalent  degrees of freedom, enabling one to focus exclusively on the interactions of multivalent ions and the fixed macromolecular surface charges on the leading {\em single-particle} level obtained through virial expansion. The single-particle characteristics of strong-coupling phenomena, arising due to multivalent ions  near charged surfaces \cite{hoda_review,Naji_PhysicaA}, and the collective, mean-field, characteristics of monovalent ions, producing Debye screening effects, are thus both captured within a single framework.

The Green's function associated with electrostatic interactions in the described model, $G(\mathbf{r},\mathbf{r}')$, representing the effective interaction potential of two test unit charges placed at positions $\mathbf{r}$ and $\mathbf{r}'$ outside the metallic core (centered at the origin), can be derived as (see the Appendices)
\begin{widetext}
\begin{equation}
G(\mathbf{r},\mathbf{r}')=\frac{\rme^{-\kappa|\mathbf{r}-\mathbf{r}'|}}{4\pi\varepsilon\varepsilon_0|\mathbf{r}-\mathbf{r}'|}+\frac{\kappa^2R_0\,\rme^{2\kappa R_0}}{4\pi\varepsilon\varepsilon_0 (1+\kappa R_0)}k_0(\kappa r)k_0(\kappa r')-\frac{\kappa}{4\pi\varepsilon\varepsilon_0}\sum_{l=0}^{\infty}(2l+1)\frac{i_l(\kappa R_0)}{k_l(\kappa R_0)}k_l(\kappa r')k_l(\kappa r)P_l(\cos\vartheta),
\label{eq:green_func_main}
\end{equation}
\end{widetext}
where  $i_l(\cdot)$ and $k_l(\cdot)$ are modified spherical Bessel functions of the first and second kind, respectively, $P_l(\cdot)$ are Legendre polynomials, and we have defined $r=|\mathbf{r}|$, $r'=|\mathbf{r}'|$, and  $\vartheta$ as the angle between $\mathbf{r}$ and $\mathbf{r}'$. The first term in Eq.  \eqref{eq:green_func_main} incorporates the direct screened Coulomb interaction between the test charges,  with the inverse screening length $\kappa$ defined through $\kappa^2=4\pi \ell_{\mathrm{B}}(2n_0+|q|c_0)$, where $\ell_{\mathrm{B}}=e_0^2/(4\pi\varepsilon\varepsilon_0k_{\mathrm{B}}T)$ is the Bjerrum length and  $2n_0+|q|c_0$ is the total bulk concentration of monovalent ions. The second and third terms in Eq.  \eqref{eq:green_func_main}  together give the  contributions from the polarization effects. These contributions stem from (i) the induced polarization (``dielectric image'') charges produced by the test charges within the metallic  NP core, and (ii) the induced polarization (``salt image'') charges produced because of the exclusion of the surrounding polarizable (and globally neutral), monovalent, ionic solution from the core region; the latter would be absent if the monovalent screening solution was present everywhere. It is however important to note that these two types of polarization effects are intrinsically entangled and enter in both the second and the third terms; also that the image charges  in the metallic core cannot in general be conceived as individual Kelvin images \cite{SCdressed1,SCdressed2,SCdressed3}.

\subsection{Configurational Hamiltonian and pressure}
\label{subsec:H_P}

The Green's function \eqref{eq:green_func_main} can be used to construct the configurational Hamiltonian of any given arrangements of $N$ multivalent ions (labeled by subscripts $i, j=1,\ldots,N$ for their positions $\{\mathbf{r}_i\}$) as
\begin{equation}
\label{eq:Hamiltonian_main}
H=\sum_{i=1}^{N} U_{ii} +\sum_{i>j=1}^{N} U_{ij} + \sum_{i=1}^N U^{\sigma}_i + U^{\sigma\sigma},
\end{equation}
where $U_{ii}$ is the self-energy of the $i$th multivalent ion, giving its self-interaction with its own image charges; $U_{ij}$ is the contribution due to direct screened Coulomb interaction between distinct multivalent ions $i$ and $j$ and the cross interactions between these ions and their respective image charges; $U^{\sigma}_i$ is the contribution due to interactions between the $i$th multivalent ion $i$ and the two charged (inner and outer) shells, subscripted by $\alpha, \beta=1, 2$  for their radius and surface charge density; and $U^{\sigma\sigma}$ is the contribution due to the interaction between the two charged shells. The image contributions to the ion-shell and shell-shell interactions are systematically included in  $U^{\sigma}_i$ and $U^{\sigma\sigma}$.  These contributions can explicitly be calculated as (see the Appendices)
\begin{widetext}
\begin{eqnarray}
&&\frac{U_{ii}}{k_{\mathrm{B}}T}= -\frac{1}{2} q^2\ell_{\mathrm{B}}\kappa\sum_{l=0}^\infty(2l+1)\frac{i_l(\kappa R_0)}{k_l(\kappa R_0)}k_l^2(\kappa r_i)+ q^2\ell_{\mathrm{B}}\frac{\kappa^2 R_0\,\rme^{2\kappa R_0}}{2(1+\kappa R_0)}k_0^2(\kappa r_i),
\label{eq:U_ii}
\\
&&\frac{U_{ij}}{k_{\mathrm{B}}T}= q^2\ell_{\mathrm{B}}\frac{\rme^{-\kappa|\mathbf{r}_i-\mathbf{r}_j|}}{|\mathbf{r}_i-\mathbf{r}_j|}+ q^2\ell_{\mathrm{B}}\frac{\kappa^2 R_0 \,\rme^{2\kappa R_0}}{(1+\kappa R_0)}k_0(\kappa r_i)k_0(\kappa r_j)- q^2\ell_{\mathrm{B}}\kappa \sum_{l=0}^\infty(2l+1)\frac{i_l(\kappa R_0)}{k_l(\kappa R_0)}k_l(\kappa r_i)k_l(\kappa r_j)P_l(\cos\vartheta_{ij}),
\label{eq:U_ij}
\\
&&\frac{U^{\sigma}_i}{k_{\mathrm{B}}T} = \frac{2\pi q \ell_{\mathrm{B}}}{\kappa r_i}\sum_{\alpha=1}^2\sigma_\alpha R_\alpha\left[\left(\rme^{-\kappa|r_i-R_\alpha|}-\rme^{-\kappa(r_i+R_\alpha)}\right)+ \rme^{-\kappa(r_i+R_\alpha)}\left(1+\rme^{2\kappa R_0}\frac{\kappa R_0-1}{\kappa R_0+1}\right)\right],
\label{eq:U_sigi}
\\
&&\frac{U^{\sigma\sigma}}{k_{\mathrm{B}}T}=\frac{4\pi^2\ell_{\mathrm{B}}}{\kappa}\sum_{\alpha,\beta=1}^2\sigma_\alpha\sigma_\beta R_\alpha R_\beta\left[\left(\rme^{-\kappa|R_\alpha-R_\beta|}-\rme^{-\kappa|R_\alpha+R_\beta|}\right)+ \rme^{-\kappa(R_\alpha+R_\beta)}\left(1+\rme^{2\kappa R_0}\frac{\kappa R_0-1}{\kappa R_0+1}\right)\right], 
\label{eq:U_sigsig}
\end{eqnarray}
\end{widetext}
where  $r_i=|\mathbf{r}_i|$, $r_j=|\mathbf{r}_j|$, and  $\vartheta_{ij}$ is the angle between $\mathbf{r}_i$ and $\mathbf{r}_j$.

In what follows, we shall focus on the effective electrostatic or osmotic pressure, $P$, acting on the virus-like (outer) shell due to the combined effect of screened Coulomb interactions between fixed charges on the outer and inner shells and their interactions with the multivalent ions,  as expressed in the configurational Hamiltonian, Eq.  \eqref{eq:Hamiltonian_main}. We shall further examine how the dielectric image charges due to the metallic core influence the net pressure on the virus-like (outer) shell. This latter quantity can be computed from our simulations (see below) using the relation 
\begin{equation}
P = -\left(\frac{\partial \overline{H}}{\partial V_2}\right)_{\!Q_2}, 
\end{equation}
where the bar represents the numerically evaluated (thermal) average over different equilibrium configurations of multivalent ions,  $V_2=4\pi R_2^3/3$ is the outer shell volume, and the partial derivatives are taken at fixed value of the total surface charge of the shell,  ${Q_2}=4\pi R_2^2 \sigma_2$. The net pressure on the outer shell can be written as
\begin{equation}
P=P_{DH}+P_q, 
\end{equation}
where $P_{DH}$ is the baseline pressure acting on the outer shell in the absence of multivalent ions and $P_q$, on the other hand, gives the pressure contribution explicitly stemming from the interactions of the shells with the multivalent ions, fully incorporating the interactions due to their respective salt/dielectric images produced by the metallic core (thus, $P_q=0$ when $q=0$). The baseline pressure can be written as 
\begin{equation}
\label{eq:P_DH_decompose}
P_{DH}=P_{\sigma_2\sigma_2}+P_{\sigma_2\sigma_1}, 
\end{equation}
where $P_{\sigma_2\sigma_2}>0$ is the pressure component arising from the self-energy of the outer shell and $P_{\sigma_2\sigma_1}$ is the pressure component arising from the interaction of the outer shell charge with the fixed charge on the inner shell, both systematically accounting for salt/dielectric image effects produced by the metallic core. The two can be obtained explicitly as (see the Appendices)
\begin{widetext}
\begin{eqnarray}
\label{eq:PDH_1}
&&P_{\sigma_2\sigma_2}=\frac{\sigma_2^2}{2\varepsilon \varepsilon_0} \left\{ \frac{1}{\kappa R_2}+\rme^{-2\kappa (R_2-R_0)}\left(\frac{\kappa R_0-1}{\kappa R_0+1}\right)\left(1+\frac{1}{\kappa R_2}\right) \right\}, 
\\
\label{eq:PDH_2}
&&P_{\sigma_2\sigma_1}=\frac{\sigma_2\sigma_1}{2\varepsilon \varepsilon_0} \frac{R_1}{R_2}\left\{ \rme^{-\kappa (R_2-R_1)}+\rme^{-\kappa (R_1+R_2-2R_0)}\left(\frac{\kappa R_0-1}{\kappa R_0+1}\right) \right\}\left(1+\frac{1}{\kappa R_2}\right).
\end{eqnarray}
\end{widetext}
The  multivalent-ion pressure $P_q$ can be written as 
\begin{equation}
P_q=-\frac{q\sigma_2}{8\pi\varepsilon \varepsilon_0\kappa R_2^2}\sum_{i=1}^N \overline{\Pi(\mathbf{r}_i)}
\,,
\end{equation}
where we have (see the Appendices)
\begin{eqnarray}
\label{eq:P-q}
&&{\Pi}(\mathbf{r}_i) \!= \frac{\rme^{-\kappa|r_i-R_2|}}{r_i}\!\left(\kappa R_2\,{\mathrm{sgn}}(r_i-R_2)\!-1\right)\!\nonumber\\
&&-\frac{\rme^{-\kappa (r_i+R_2-2R_0)}}{r_i}\times(1+\kappa R_2)\left(\frac{\kappa R_0-1}{\kappa R_0+1}\right),
\end{eqnarray}
where ${\mathrm{sgn}}(\cdot)$ is the sign function. 

We shall also discuss later the reference case of a model VLP with no  NP core. This case helps discern the effects of image charges produced by the  NP core from other factors. The corresponding expression for $P_{DH}$ and $P_q$ in this latter case 
are given in the Appendices. 

\subsection{Simulation details}

The configurational Hamiltonian, Eq. \eqref{eq:Hamiltonian_main}, can be used with appropriately designed Monte Carlo (MC) simulations to calculate equilibrium properties of the considered model VLP immersed in an asymmetric ionic mixture. We use the iterative canonical MC algorithm introduced by us in Refs. \cite{leili1,leili2}, by placing the VLP at the center of a cubic simulation box with periodic boundary conditions. Finite system size effects are efficiently eliminated within the simulation error bars by choosing a large enough box, here of lateral size $2(R_2+8\kappa^{-1})$. The iterative method enables producing the bulk conditions in a few Debye screening distances from the outer shell  and with given bulk ionic concentrations using prescribed values of $c_0$ and $\kappa$ \cite{leili1}. The summations in Eqs.  \eqref{eq:U_ii}-\eqref{eq:U_sigsig} are calculated by choosing a series cutoff guaranteeing a relative truncation error  of $10^{-7}$ or smaller. In the simulations, the series summations are first tabulated by taking a mesh in the coordinates space such that the maximum error generated in the calculation of the energy (per  $k_{\mathrm{B}}T$) is of the order $10^{-3}$. The energy for arbitrary configurations of multivalent ions is computed using tricubic interpolation methods \cite{TricubicPaper}.  The relative error in reproducing the reported bulk concentration is of the order 0.1\%, which is smaller than the relative (sampling) error bar of 1-2\% consistently obtained from our simulations using the block-averaging methods \cite{BlockAvrg}. The simulations run for at least $6\times 10^7$ MC steps per particle with the first $10^7$ steps used for equilibration purposes.

\subsection{Choice of parameter values}
\label{subsec:parameters}

Our model parameters include the radius $R_0$ of the NP core, the inner and outer shell radii $R_1$ and $R_2$ and their surface charge densities $\sigma_1$ and $\sigma_2$, respectively, the multivalent-ion charge valency $q$, the inverse screening length $\kappa$ (supplemented by the ionic densities $n_0$ and $c_0$), and the solvent dielectric constant, which we shall equate to that of water at room temperature, $\varepsilon = 80$. Our primary goal is to investigate the generic electrostatic properties of the model VLP with a metallic core in the presence of (positively and negatively charged) multivalent ions and, as such, structural details that may be present in actual experimental systems are largely ignored (see Section \ref{sec:conclusion}). In order to bring out the salient roles of image charges and multivalent ions, especially in the effective pressure they produce on the capsid (outer shell), we explore typical regions of the parameter space by fixing the values of system parameters that are of less immediacy to our analysis, i.e., the radii of the NP core and the two charged shells, the charge density of the capsid $\sigma_2$, and vary the other parameters.

In experiments with plant viruses (such as BMV, RCNMV, or CCMV) encapsidating a single coated gold NP~\cite{DragneaPNAS2007,DragneaACSNano2010,DragneaNanoLett2006,Loo2006,Loo2007,Aniagyei2009}, the NP core radius typically ranges from 3-12~nm with a coating layer of thickness around $2$~nm~\cite{DragneaACSNano2010}. The size of the assembled viral capsid naturally corresponds with the size of the core; for instance, a sufficiently large core will lead to the assembly of wild-type BMV with the capsid triangulation number ${\mathcal T} = 3$, while a smaller core will lead to an assembly of a smaller ${\mathcal T} = 1$ capsid \cite{DragneaPNAS2007}.
To provide an estimate of typical surface charge densities on the assembled capsids, we again take the example of BMV capsid \cite{DragneaACSNano2010}, which has a typical total charge of $Q=540 {\mathcal T}e_0$. 
For smaller (${\mathcal T} = 1$) capsids as, for instance, obtained by NP cores of radius $R_0=3$~nm \cite{DragneaPNAS2007}, one can take the typical values of $R_1=5$~nm \cite{DragneaACSNano2010} and $R_2=6$~nm \cite{DragneaNanoLett2006}, 
giving an effective surface charge density of $\sigma_2\simeq 1.2$~$\mathrm{nm}^{-2}$ for the outer shell. 
Without loss of generality, we fix these typical numerical values for the most part in our numerical simulations and vary $\sigma_1$, $\kappa$ and, in some cases, $R_1-R_0$.

The surface charge density of the coating layer,  $\sigma_1$, is varied within the range $\sigma_1\in[0,5]$~$\mathrm{nm}^{-2}$, which is consistent with the values used or estimated in previous studies \cite{vdSchoot2015,DragneaACSNano2010}.
The inverse screening length is varied in the range $\kappa=0.5-2$~$\mathrm{nm}^{-1}$  for three different cases of multivalent ion valencies, $q=+2$, $q=+4$ and $q=-4$. We fix multivalent ion concentration  as $c_0=5$~$\mathrm{mM}$ for divalent cations, and  $c_0=1$~$\mathrm{mM}$ for tetravalent cations and anions. Hence, for instance, the numerical value of $\kappa= 0.78$~$\mathrm{nm}^{-1}$ corresponds  to monovalent ion concentration of $n_0\simeq 53$~$\mathrm{mM}$ for divalent cations, and  $n_0\simeq56$~$\mathrm{mM}$  for tetravalent cations/anions. These numerical values also fall consistently within the experimentally accessible ranges of values \cite{DragneaACSNano2010}.  In the simulations, we take a finite radius of $a=0.15~\mathrm{nm}$ for the multivalent ions. Needless to say that our models has an obvious symmetry with respect to sign inversions $q\rightarrow-q$ and $\sigma_{1,2}\rightarrow -\sigma_{1,2}$. 

Though some of the numerical values listed above are adopted directly from the BMV experiments, other ranges of parameter values corresponding to other NPs and other viral capsids (such as CCMV) \cite{PSS_CCMV} can be equally well addressed with our generic electrostatic model for VLPs. 

\begin{figure}[t!]
\includegraphics[width=6cm]{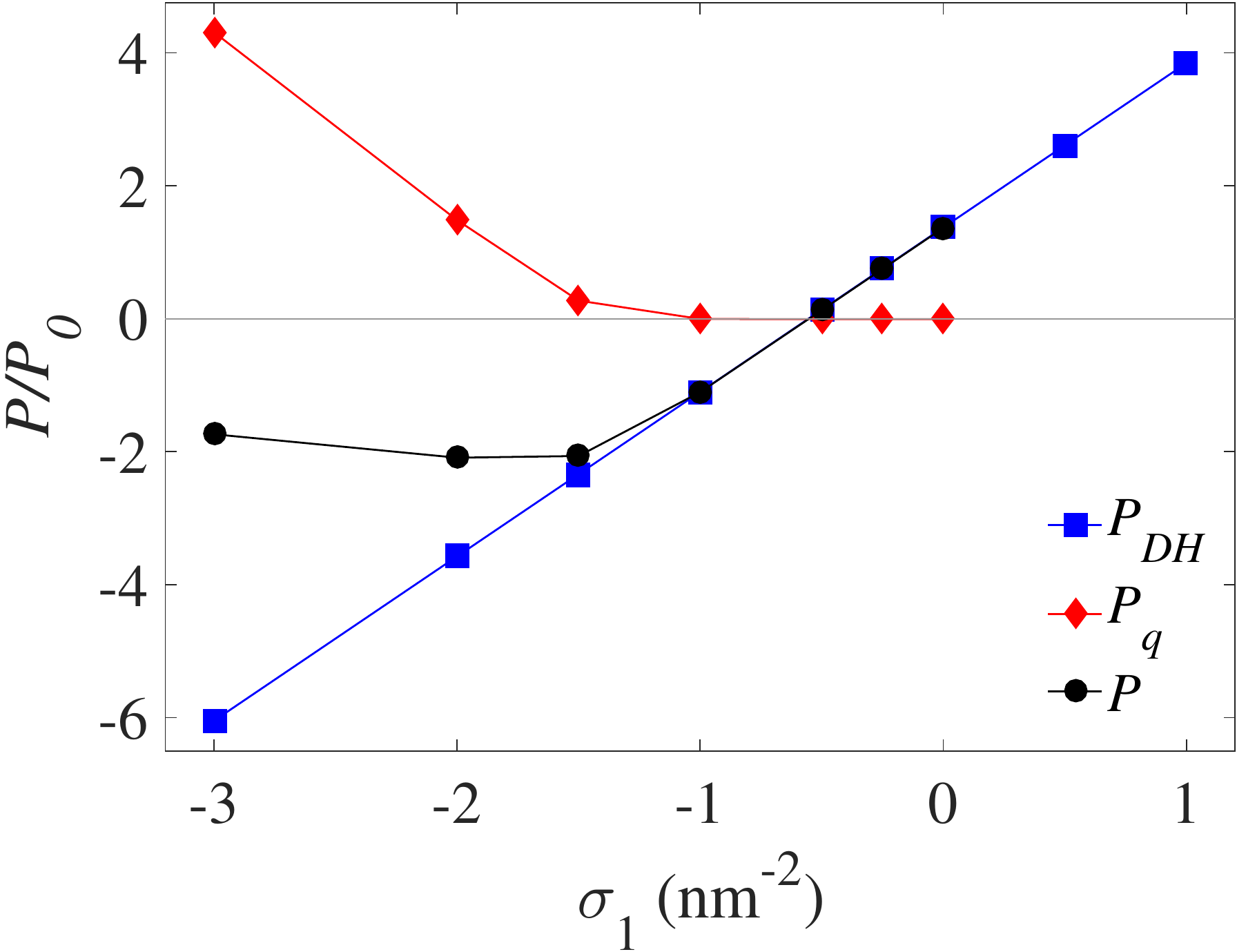}
\caption{Rescaled net osmotic pressure, $P$, acting on outer shell of the model VLP, and its different components, $P_{DH}$ and $P_q$, plotted as a function of the surface charge density of the NP coating layer for divalent cations ($q=+2$), the outer shell surface charge density $\sigma_2=+1.2$~$\mathrm{nm}^{-2}$ and radius $R_2= 6$~nm, the NP radius $R_0 = 3$~nm, the coating layer thickness 2~nm ($R_1= 5$~nm), and the inverse Debye screening length $\kappa\simeq 0.78$~$\mathrm{nm}^{-1}$. We have defined $P_0=1 k_{\mathrm{B}}T/{\mathrm{nm}}^{3} \simeq 41$~atm.}
\label{fig:AllPs_q2}
\end{figure}

\section{Results and discussion}

\subsection{Pressure components: Cations vs anions}

In Fig. \ref{fig:AllPs_q2}, we show our simulation results for the net osmotic pressure $P$ on the outer virus-like shell,  and its components $P_{DH}$ and $P_q$, as defined in Section \ref{subsec:H_P},  as a function of the coating surface charge density of the metallic NP core (or the inner shell charge density), $\sigma_1$. Here, we have used  divalent cations ($q=2$) with the typical choices of values for fixed parameters as noted in Section \ref{subsec:parameters}; i.e., $\sigma_2=1.2$~$\mathrm{nm}^{-2}$, $\kappa= 0.78$~$\mathrm{nm}^{-1}$ (with fixed $c_0=5$~mM), $R_0 = 3$~nm, $R_1= 5$~nm and  $R_2= 6$~nm. In the plots hereafter we rescaled the pressure values with $P_0=1 k_{\mathrm{B}}T/{\mathrm{nm}}^{3} \simeq 41$~atm. 

As seen in the figure, the baseline pressure $P_{DH}$ (blue squares) acting on the outer shell in the absence of multivalent ions increases linearly with $\sigma_1$, which is in accordance with Eq. \eqref{eq:PDH_2}. $P_{DH}$ changes sign from negative (inward-directed pressure on the outer shell) to positive values (outward-directed pressure) as $\sigma_1$ is increased above $\sigma_1^\ast\simeq -0.55$~$\mathrm{nm}^{-2}$. The  change of sign in $P_{DH}$ occurs at a negative value of $\sigma_1$ because of our choice of a positive outer shell charge density $\sigma_2>0$; that is, the  positive self-pressure of the outer shell $P_{\sigma_2\sigma_2}$ can be balanced by the negative inter-shell pressure component $P_{\sigma_2\sigma_1}$ (giving the inward-directed pull of the outer shell by the inner shell) only if the two shells are oppositely charged; see also Eq. \eqref{eq:P_DH_decompose}. 

Another point to be noted is that the pressure component due to divalent cations, $P_q$ (red diamonds), vanishes for $\sigma_1>\sigma_1^{\ast\ast}\simeq -1$~$\mathrm{nm}^{-2}$, where the net pressure (black circles) equals the base pressure, $P\simeq P_{DH}$.  $P_q$ is nonzero only for $\sigma_1<\sigma_1^{\ast\ast}\simeq -1$~$\mathrm{nm}^{-2}$, where it takes sizably large (outward-directed pressure) values, partially balancing the negative (inward-directed) pressure due to $P_{DH}$, or more accurately, $P_{\sigma_2\sigma_1}$; hence, producing a non-monotonic behavior in the net pressure, $P$, as a function of $\sigma_1$ (Fig.  \ref{fig:AllPs_q2}). The reported net pressure in this latter regime of coating surface charge densities is thus a direct consequence of inter-shell attraction $P_{\sigma_2\sigma_1}$, giving net values of around $P\simeq -40$ to $-80$~atm in actual units. This implies an electrostatically favorable situation for the formation and stability of the resulting VLP for $\sigma_1<\sigma_1^{\ast\ast}$, in contrast to the opposite scenario, which is predicted to hold in the regime $\sigma_1> \sigma_1^\ast$, where the net pressure becomes positive. The conclusion that a minimum negative value of $|\sigma_1|$ is required to enable formation of stable VLPs with divalent cations is generally consistent with recent findings in the BMV context \cite{DragneaACSNano2010}, where the BMV encapsidation of a coated gold NP is reported to occur only for  $\sigma_1<-2$~$\mathrm{nm}^{-2}$, roughly corresponding to the location of the minimum net pressure (black circles) in Fig. \ref{fig:AllPs_q2}.

\begin{figure}[t!]
\includegraphics[width=8.5cm]{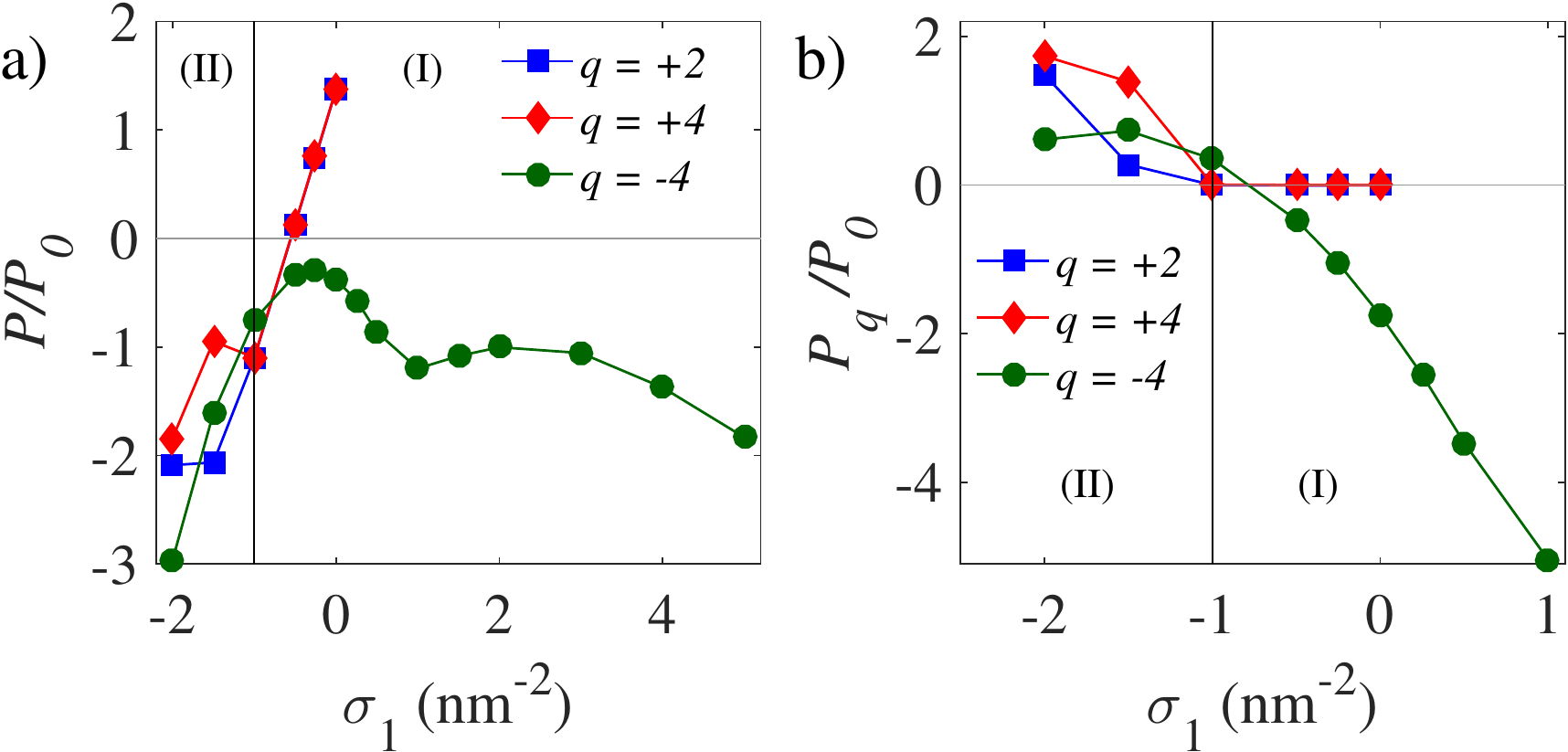}
\caption{Same as Fig. \ref{fig:AllPs_q2} but shown are (a) the net osmotic pressure, and (b) the pressure contribution stemming from multivalent ions, $P_q$, for three different values of $q$. The vertical line shows the coating layer charge density $\sigma_1^{\ast\ast}$ as introduced in the text. 
}
\label{fig:Allq_Ps}
\end{figure}

The net pressure on the outer shell shows qualitatively similar behavior for varying cation charge valency (compare $q=2$ and 4 in Fig. \ref{fig:Allq_Ps}a) with significant deviations occurring only at sufficiently large  magnitudes of (negative) $\sigma_1$. However, a remarkably different behavior is observed if we use anionic multivalent ions, e.g., $q=-4$ (Fig. \ref{fig:Allq_Ps}a). This is an interesting case as, in contrast to the case of cations,  multivalent anions are electrostatically repelled from the inner NP coating layer ($\sigma_1<0$), while they are attracted more strongly to the outer shell ($\sigma_2>0$) as we shall discuss later in this section. The net pressure on the outer shell, $P$, becomes negative for multivalent anions (black circles in Fig. \ref{fig:Allq_Ps}a) across the whole range of $\sigma_1$ plotted in the figure. Thus, while in region (II) in the figure ($\sigma_1< \sigma_1^{\ast\ast}$), one can obtain negative net pressure in both cases of multivalent cations and multivalent anions (even though $P_q$ remains positive in either case in region (II); see Fig. \ref{fig:Allq_Ps}b), only multivalent anions can produce a negative net  pressure in region (I) ($\sigma_1> \sigma_1^{\ast\ast}$). A more detailed comparison between the two exemplary cases with $q=2$ and $q=-4$ ions will be given in Sections \ref{subsec:images} and \ref{subsec:empty}. 
Here, it is important to note that different mechanisms are at work in regions I and II, and also for cationic vs anionic multivalent ions. While, as noted above, the resulting negative net pressure in region (II) can be understood as indicating a dominant inter-shell attraction component, $P_{\sigma_2\sigma_1}$, regardless of the multivalent ion charge, the dependence of the net pressure on $q$ in region (I) can be elucidated by examining the accumulation of multivalent ions within the VLP. 

\begin{figure}[t!]
\includegraphics[width=6cm]{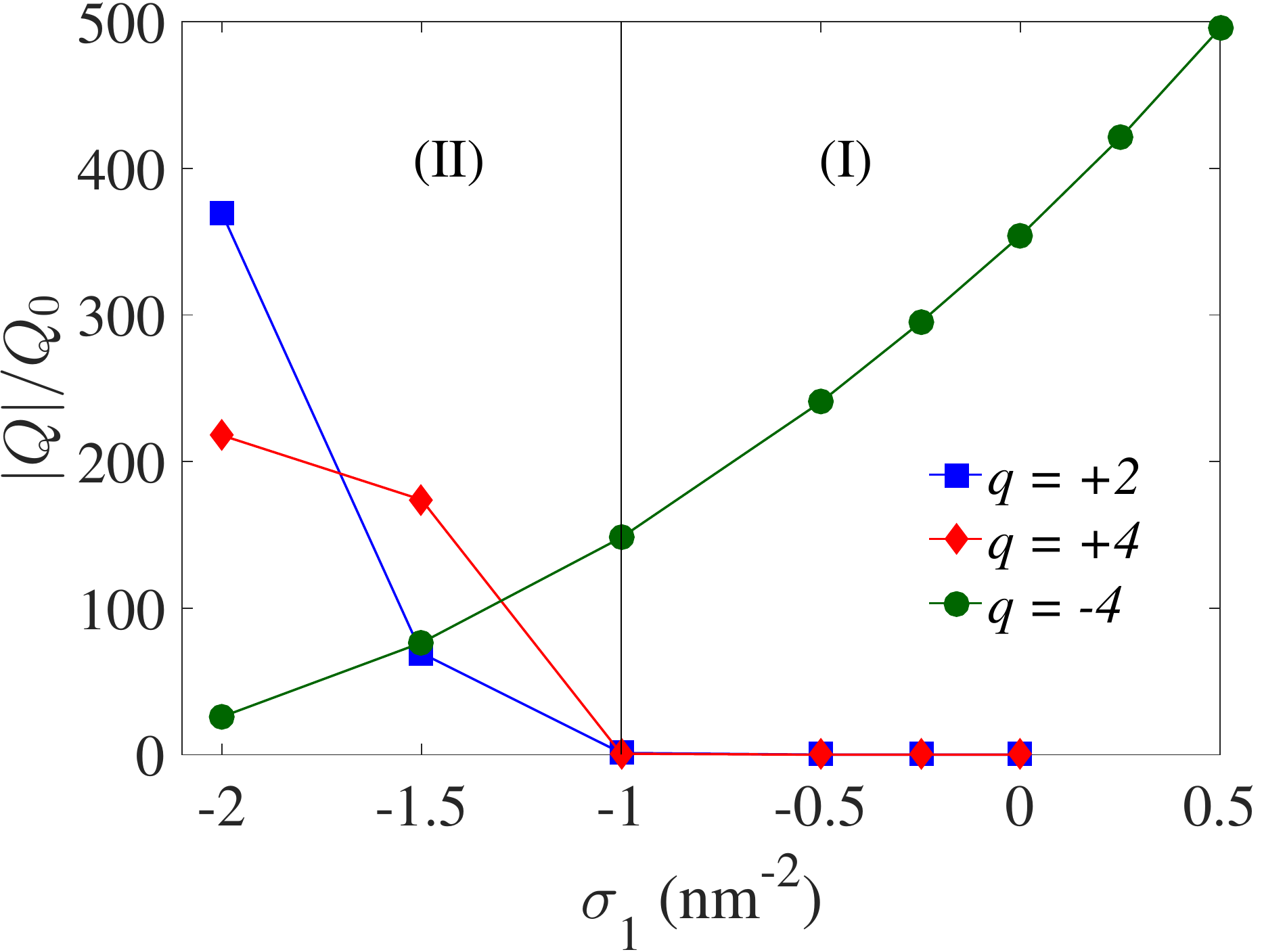}
\caption{Absolute value of the total multivalent-ion charge, $|Q|$, accumulated within  inter-shell space of the VLP, $R_1<r<R_2$, plotted as a function of the surface charge density of the NP coating layer, for three different values of $q$. The results are rescaled with the characteristic charge $Q_0=(4\pi/3)(R_2^3-R_1^3)|q|e_0c_0$. See Fig. \ref{fig:AllPs_q2} for other parameters. 
}
\label{fig:Q}
\end{figure}

\subsection{Multivalent-ion accumulation within VLP}

\begin{figure}[t!]
\includegraphics[width=6cm]{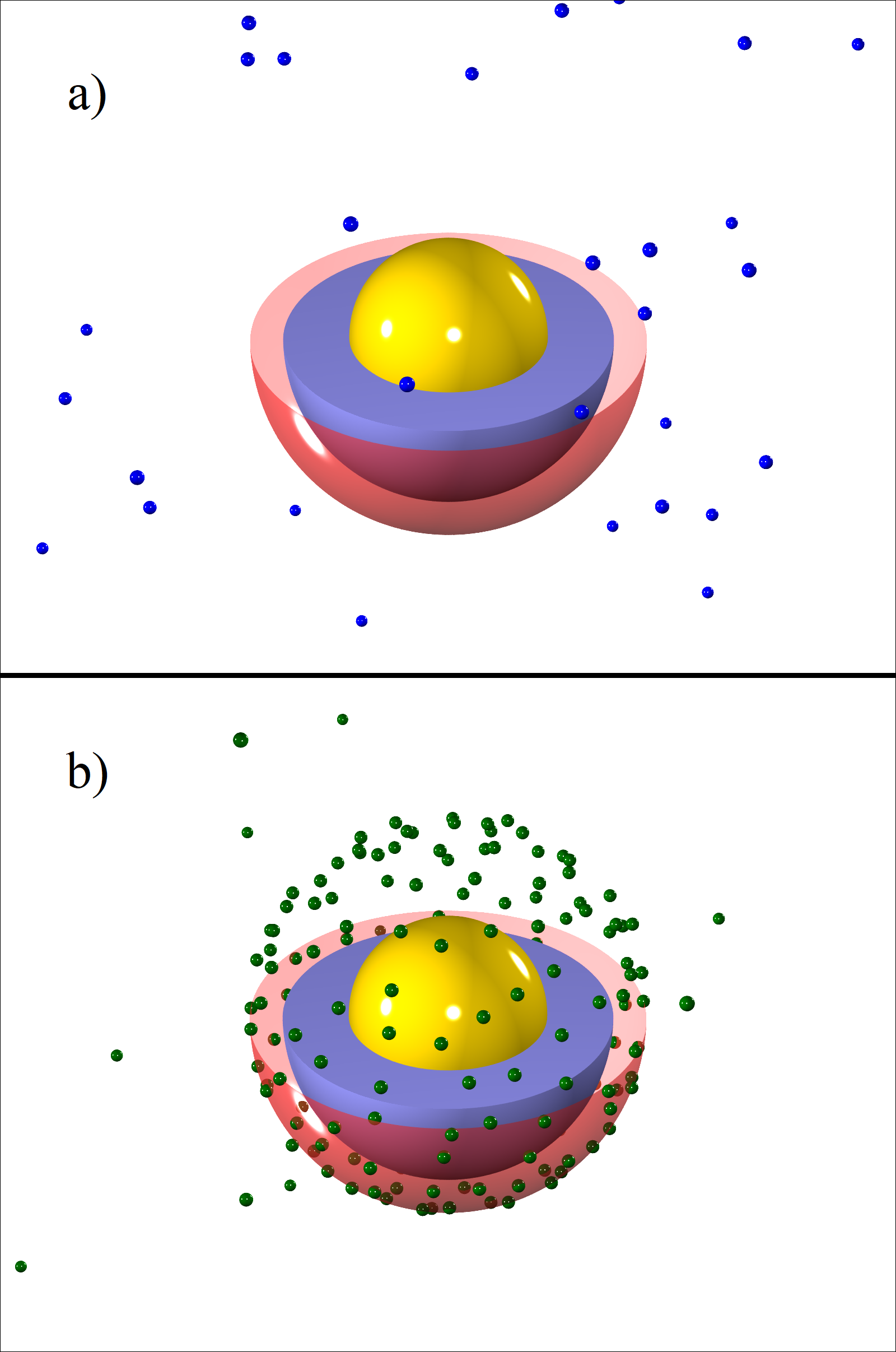}
\caption{Simulation snapshot showing positions of multivalent ions  in and outside the model VLP for (a)  divalent cations (shown as small blue spheres) and (b) tetravalent anions (shown as small green spheres) for fixed  $\sigma_1=0$ and other parameter values given in Fig. \ref{fig:AllPs_q2}. For the sake of illustration, the upper half of the outer shell (shown schematically as a red transparent sphere) and the NP coating layer (shown as a thick blue shell) are removed in the picture, while the core metallic NP is shown in whole (yellow sphere). }
\label{fig:snapshot}
\end{figure}

The absolute value of the total multivalent-ion charge, $|Q|$, accumulated within  inter-shell space of the VLP, $R_1<r<R_2$, can directly be measured from our simulations. We rescale the accumulated charge with the characteristic charge $Q_0=(4\pi/3)(R_2^3-R_1^3)|q|e_0c_0$. The results, plotted in Fig. \ref{fig:Q} as a function of $\sigma_1$, clearly show that, while multivalent anions exhibit a smoothly decreasing degree of accumulation within the VLP by decreasing $\sigma_1$ from positive to negative values, multivalent cations exhibit a complete depletion from within the VLP in the parameter region (I) (giving $|Q|\simeq 0$), followed by a sharp increase in $|Q|$ as one enters region (II) by decreasing $\sigma_1$ (see also simulation snapshots in Fig. \ref{fig:snapshot}). Although these behaviors can generally be understood based on the net charge of the VLP, i.e., the sum of the inner and outer shell charges, becoming positive (negative) in region (I) (region (II)), the detailed behavior of $Q$ is determined by the combined effect of the three interaction terms \eqref{eq:U_ii}-\eqref{eq:U_sigi} and vary depending on the precise choice of system parameters. 

The behavior of accumulated multivalent-ion charge in Fig. \ref{fig:Q} also provides a better insight into the behavior of $P_q$ as shown in Fig. \ref{fig:Allq_Ps}b. The near complete depletion of multivalent cations from within the VLP in region (I) clearly explains why $P_q\simeq 0$ and, hence, $P\simeq P_{DH}$, in the mentioned parameter region. Also, the stronger accumulation of multivalent cations within the VLP by decreasing $\sigma_1$ in region (II) (making $\sigma_1$ more negative; Fig. \ref{fig:Q}) shows that the positive $P_q$ (blue and red symbols in Fig. \ref{fig:Allq_Ps}b) results from the stronger repulsion these ions impart on the outer shell, which is also positively charged  (formally, this repulsion is embedded in Eq. \eqref{eq:U_sigi}). 

In the case of multivalent anions, the positive $P_q$ in region (II) (black symbols in Fig. \ref{fig:Allq_Ps}b) results from the image charges of these ions in the metallic NP core that will be positively charged; hence, producing the only source of repulsion on the outer shell (as embedded in Eq. \eqref{eq:U_ii}) that may stem from the multivalent ions in this case. Another distinct aspect of multivalent anions is that the pressure component $P_q$ changes sign to take negative values of large magnitude in region (I), engendering also a large negative net pressure $P$ on the outer shell, as noted before (Figs. \ref{fig:Allq_Ps}a). We shall return to the underlying cause of this large negative pressure in Section \ref{subsec:empty}. 

We conclude this section by emphasizing that the change of sign in the pressure component $P_q$ in going from region (II) to region (I) is essential in maintaining a negative osmotic pressure on the outer shell across a broad range of values for $\sigma_1$ in the case of multivalent anions. Our results thus also suggest that multivalent anions present a more robust case (in contrast to multivalent cations) in stabilizing the VLP, irrespective of the sign and magnitude of the surface charge density of the NP coating layer.  

\subsection{Image charge effects}
\label{subsec:images}

The effects due to the image charges on the net pressure, $P$, can be assessed by comparing the results obtained in the case of a VLP containing a metallic NP with those obtained in an equivalent model in which the dielectric constant of the NP core is set equal to that of the aqueous solution, $\varepsilon=80$, while the coating charge density, $\sigma_1$,  is kept fixed and the monovalent salt (screening) ions are allowed to permeate within the core region ($r<R_0$). These cases are labeled  in the plots by ``NP'' and ``no NP'', respectively. Note that, in the latter (``no NP'') case, both the  dielectric images and also the much weaker salt images are eliminated \cite{SCdressed1,SCdressed2,SCdressed3,perspective}.  (In the simulations of the ``no NP'' case, the image elimination is done by using only the free-space part of the Green's function, i.e., by keeping the first term of Eq. \eqref{eq:green_func_main} and omitting the other two terms.) 

\begin{figure}[t!]
\includegraphics[width=8.5cm]{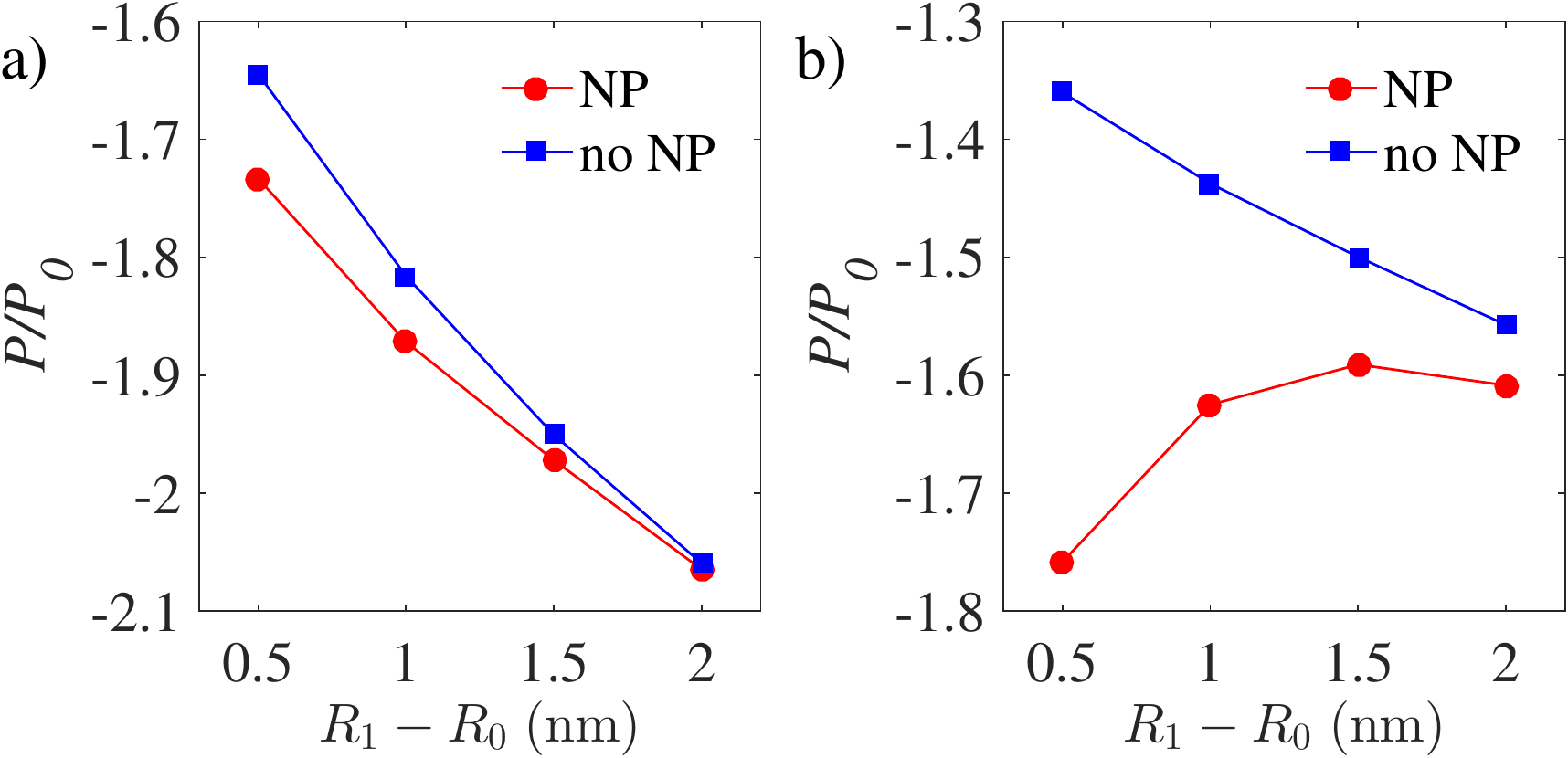}
\caption{Rescaled net osmotic pressure is plotted as a function the coating layer thickness, $R_1-R_0$, for (a) $q = +2$ and (b) $q = -4$, and for fixed  $\sigma_1=-1.5$~$\mathrm{nm}^{-2}$. Here, we have $R_2-R_1=1$~nm.  Other parameters are as in Fig. \ref{fig:AllPs_q2}.}
\label{fig:Coat_Thick}
\end{figure}

The image charge effects turn out to be small for divalent cations ($q=2$) for the parameters chosen in Fig. \ref{fig:Coat_Thick}a, while they turn out to be significant for tetravalent anions ($q=-4$) as shown in Fig. \ref{fig:Coat_Thick}b. In the former case, the net pressure on the outer shell, shown as a function of the coating layer thickness, $R_1-R_0$, in the figure, displays a relative change of only a few percent (around 5\%) between the two cases with and without an ideally polarizable metallic core (``NP'' vs ``no NP'').  In the case of multivalent anions, the effects of image charges are more drastic, changing the net-pressure profile from a monotonic to a non-monotonic one as shown in Fig. \ref{fig:Coat_Thick}b. In this case, even at the largest coating layer thickness shown in the plot, $R_1-R_0=2$~nm, the pressure decrease due to the  image charges turns out to be about 2~atm in actual units, indicating that, in the process of encapsidating a metallic core, VLP stabilization can more significantly be assisted by the images of multivalent anions than those of multivalent cations. In both cases, the  difference between ``no NP'' and ``NP'' cases decreases as $R_1-R_0$ becomes comparable to or larger than the Debye screening length (here, $\kappa^{-1}\simeq 1.28$~nm).

To understand the difference in the image-charge effects found in the two cases mentioned above, one should first note that inserting (removing) the metallic NP core has the effect of decreasing (increasing) the net pressure, or making it more attractive (repulsive), by itself. This is due to the fact that, in the presence of a metallic core, the outer shell charge distribution, $\sigma_2$, is pulled inward by its own oppositely charged image, being produced in the core (formally, this effect is embedded in Eq. \eqref{eq:U_sigsig}). This decrease in the net pressure, $P$, is independent of the choice of $q$ and occurs equally in both cases shown in Figs. \ref{fig:Coat_Thick}a and b. Therefore, the difference in the pressure drop in the two figures represents the intricate ways in which the pressure component due to multivalent ions, $P_q$, is affected by the NP insertion. Again, one can generally expect a larger accumulation of multivalent ions inside the VLP upon the core insertion, which is driven by the attraction of multivalent ions with their (oppositely charged) images in the core; therefore, based only on their direct electrostatic interactions with the positive outer-shell charge distribution, one can anticipate a more positive $P_q$ in the case of multivalent cations than in the case of multivalent anions. This rough argument however cannot explain the detailed features of the net-pressure profiles and misses other competing factors such as the additional counteracting  attractive (repulsive) pressure that the image charges of multivalent cations (anions) exert on the outer shell, and the positive image charge of the coating layer, which appears upon the NP insertion in the VLP core, producing an additional positive pressure on the outer shell. 

\begin{figure}[t!]
\includegraphics[width=6cm]{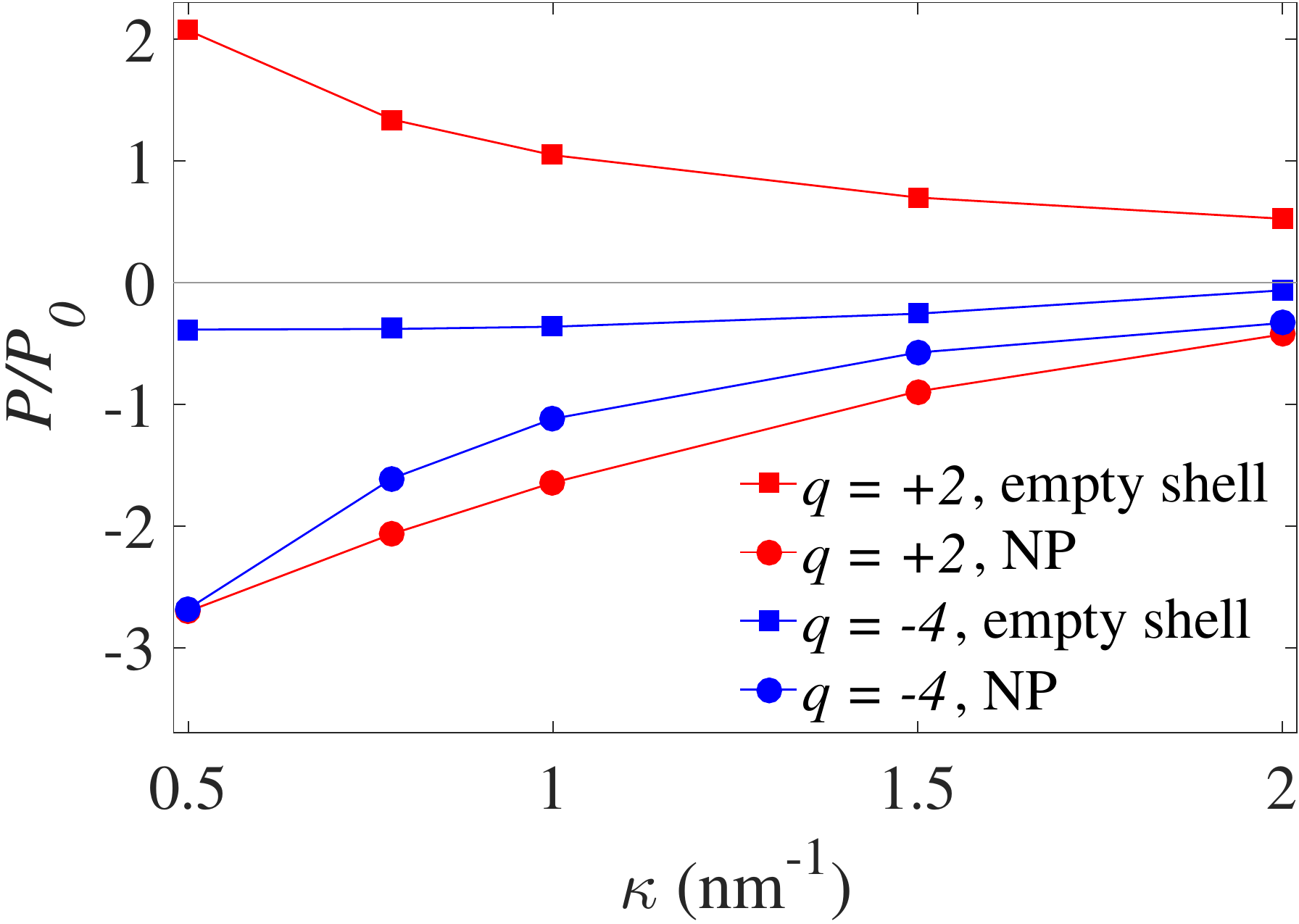}
\caption{Rescaled net osmotic pressure plotted as a function of the inverse Debye screening length, $\kappa$, for $q = +2$ and $q = -4$. We have compared the data obtained for a metallic-core VLP with those obtained for an empty shell (labelled ``NP'' and ``empty shell'' in the graph, respectively) with fixed $\sigma_1=-1.5$~$\mathrm{nm}^{-2}$. Other parameters are as in Fig. \ref{fig:AllPs_q2}.
}
\label{fig:Ptot_k}
\end{figure}

\subsection{Model VLP with and without NP core}
\label{subsec:empty}

We proceed by comparing the results for a VLP containing a metallic NP (labeled by ``NP'') with those obtained in an equivalent model in which the VLP contains no NPs, representing an ``empty shell'', whose entire inside volume is accessible to multivalent and monovalent (salt) ions.   In Fig. \ref{fig:Ptot_k}, the net pressure, $P$,  on the outer shell is plotted as a function of the inverse Debye screening parameter, $\kappa$, for $q=2$ and $q=-4$ (note that we have fixed $c_0=5$~mM for divalent cations, and $c_0=1$~mM for tetravalent anions; therefore, $\kappa=2$~$\mathrm{nm}^{-1}$ can be obtained by taking monovalent salt concentration of $n_0=376$~mM in the former case and $n_0=379$~mM in the latter case). 

For the parameters given in Fig. \ref{fig:Ptot_k}, the net pressure $P$ acting on the empty shell is positive, as it is dominated by the positive self-pressure, $P_{\sigma_2\sigma_2}$, of the shell in the case of divalent cations. By contrast, the net pressure becomes negative for tetravalent anions, due to the dominant negative $P_q$, which is produced itself by the attractive strong-coupling interactions mediated by tetravalent anions between the opposing parts of the empty shell. This latter strong-coupling mechanism as engendered by multivalent counterions has been discussed in detail elsewhere \cite{leili1}. This is also the reason why tetravalent anions produce the large negative pressure in region (I), especially for positive values of $\sigma_1$, in Fig. \ref{fig:Allq_Ps}a and b.

In both cases of $q=2$ and $q=-4$ in Fig. \ref{fig:Ptot_k}, however, the net pressure on the outer shell becomes negative when a metallic NP core is inserted in the shell. This is an important observation, highlighting the substantial effect of the metallic NP core in decreasing the net pressure on the outer shell and making the VLP complex (electrostatically) significantly more stable than an empty virus-like shell.

\section{Conclusion}
\label{sec:conclusion}

We study the role of electrostatic interactions in stabilization of a model virus-like particle (VLP) encapsidating a coated metallic nanoparticle (NP) in an asymmetric Coulomb fluid, which consists of a solution mixture of $1:1$ and $q:1$ salts. Using Monte-Carlo (MC) simulations within an effective dressed multivalent-ion model \cite{SCdressed1,SCdressed2,SCdressed3,perspective}, we compute and analyze the effective electrostatic (osmotic) pressure acting on the outer VLP shell and identify the regimes of positive (outward-directed) and negative (inward-directed) osmotic pressure in the case of multivalent cationic and anionic charge valencies, coating layer charge densities and thicknesses, and the bathing salt concentration. Without loss of generality, we assume that the outer shell of the VLP is positively charged and the sizes of the outer shell, the core particle and its coating layer are consistent with recent BMV experiments \cite{DragneaPNAS2007,DragneaACSNano2010}. We elucidate the interplay between various cooperating or competing factors in the electrostatic stability of the VLP, such as the electrostatic self-pressure of the outer shell, the interaction between the outer shell and the coating layer charge, the strong-coupling effects produced by multivalent (counter)ions, and the image charge effects that produced by the ideally polarizable metallic NP core. 

In the case of multivalent cations, inward-directed net pressure on the outer shell, stabilizing the VLP, is found to occur only for negative coating layer charge densities of sufficiently large magnitude, which is generally consistent with recent experimental observations of coated gold NP-encapsidation within BMV capsids  \cite{DragneaACSNano2010}. Multivalent anions can however generate negative net osmotic pressure on the outer shell for the whole range of positive and negative coating layer charge densities, with pressure magnitudes of the order of few tens of atm. This suggests that multivalent anions will play a more robust role in electrostatic stabilization of VLP particles. Our analyses also show that the image charge effects, resulting from the insertion of a metallic NP, can generally make the VLP more stable (by reducing the net pressure on the outer shell or even by changing the sign of the pressure from positive to negative) as compared with an equivalent situation where the NP core is removed from the VLP. While the dominant mechanism at work for multivalent cations (and/or for sufficiently negative coating layer charge densities) is the inter-shell attraction between the positively charged outer shell and the negatively charged coating layer charge, the dominant mechanism in the case of multivalent anions turns out to be the inward-directed osmotic pressure they create due to their strong coupling to the outer shell charge density. These effects have previously been addressed in detail in the case of empty shells or charge droplets \cite{leili1,leili2}. 

Our MC simulations are enabled by calculating the electrostatic interactions through the relevant Green's function of the system (see the Appendices), which consistently accounts for image charge effects (due to inhomogeneous dielectric constant and salt distribution in the system), salt screening effects and also strong electrostatic coupling effects  due to multivalent ions that go beyond the scope of usual mean-field theories. Our dressed multivalent-ion implementation thus  accounts for both the ionic screening effects due to the weakly coupled monovalent ions in the bathing solution and also the leading-order electrostatic correlations between multivalent ions and the opposite surface charges on the shell or NP coating layer (see Refs. \cite{leili1,leili2,SCdressed1,SCdressed2,SCdressed3,perspective,book,hoda_review}  for further details).

The present model is constructed based on several simplifying assumptions, which, despite their limitations for the applicability to specific systems at full extent, posses two key advantages. First, they enable us to provide a thorough investigation of electrostatic effects that usually turn out to be very challenging because of the long-ranged nature of Coulomb interactions and the combined interplay between various factors such as mobile (multivalent) ions and image charge related effects \cite{perspective,book,hoda_review}. As such, our model also helps circumventing difficulties in the computational implementation of 
the simulation within the present context. Secondly, our model can be used as a generic description for the NP encapsidation by a variety of proteinaceous shells in terms of a few basic parameters, whose numerical values can then be adopted according to the specific cases of interest. Our model can  be straightforwardly extended to include the different dielectric constants of the NP coating layer, that of the solution, and that of the NP core \cite{PRL_Coat}.
The more realistic aspects of viral capsids require more detailed modeling for the geometry and charge distribution of the capsid and the NP coating, stipulating more extensive coarse-grained or atomistic simulation techniques. Atomistic models can also help address the role of discrete nature of water molecules and its effect on the dielectric properties of the medium, especially inside the capsid. It is also worth mentioning that even models with the level of simplification we have used in this work \cite{leili1,leili2} would be able to capture the key electrostatic features of atomistic models; this is evidenced by the all-atom molecular-dynamics simulations of empty poliovirus capsids \cite{FullAtomSim}, showing that the pressure acting on these empty capsids inside the solution can be negative due to electrostatic interactions, in accordance with our previous findings \cite{leili1}. Other factors that can be addressed in future studies include the role  of specific-ion effects, detailed structure of multivalent ions as well as the charge regulation effects \cite{perspective}.

\begin{acknowledgments}
L.J. acknowledges funds from National Elites Foundation (Iran) and useful discussions with I. Tsvetkova, P. van der Schoot,  R. Bruinsma and K. Hejazi. A.N. acknowledges partial support from the Royal Society, the Royal Academy of Engineering, and the British Academy (UK) and partial funds from Iran Science Elites Federation and the Associateship Scheme of The Abdus Salam International Centre for Theoretical Physics (Trieste, Italy). 
L.J. and R.P. acknowledge partial support by the National Science Foundation Grant No. PHYS-1066293 and travel grant from the Simons Foundation through the Aspen Center for Physics. A.L.B. acknowledges the financial support from the Slovenian Research Agency (research core funding no. P1-0055).
\end{acknowledgments}


\begin{widetext}

\appendix

\section{Screened Coulomb Green's function in the presence of a metallic sphere}

The Green's function $G(\mathbf{r},\mathbf{r}')$, describing the electrostatic interactions of explicit charges within the dressed multivalent ion theory, is standardly obtained from the Debye-H\"uckel (DH) equations, governing the electrostatic potential in an electrolyte surrounding an ideally polarizable, metallic, nanoparticle (NP) of radius $R_0$ with constant surface (and interior) potential. Hence, by taking the center of coordinates at the center of the NP, we have 
\begin{equation}
\label{eq:Green}
\left\{
\begin{array}{lr}
G(\mathbf{r},\mathbf{r}')=C& ,\; r \leq R_0,\\
\nabla^2G(\mathbf{r},\mathbf{r}')-\kappa^2G(\mathbf{r},\mathbf{r}')=-\frac{\delta(\mathbf{r}-\mathbf{r}')}{\varepsilon\varepsilon_0}& ,\; r > R_0, 
\end{array}
\right.
\end{equation}
where $C$ is a constant. The solution to the above set of equations in the region outside the spherical NP can be expressed as the sum of a ``special'' solution (first term below), representing the bulk solution $
G_0(\mathbf{r},\mathbf{r}')={\rme^{-\kappa\dist}}/(4\pi\varepsilon\varepsilon_0\dist)$, and a ``homogenous'' solution (second term below) due to the presence of the NP  \cite{Arfken},
\begin{equation}
G(\mathbf{r},\mathbf{r}')=\frac{1}{4\pi\varepsilon\varepsilon_0}\frac{\exp(-\kappa|\mathbf{r}-\mathbf{r}'|)}{|\mathbf{r}-\mathbf{r}'|} +
\sum_{l=0}^{\infty}B_lk_l(\kappa r)P_l(\cos\vartheta),
\end{equation}
where  $k_l(\cdot)$ is the modified spherical Bessel function of the second kind, $P_l(\cdot)$ are Legendre polynomials, and we have defined $r=|\mathbf{r}|$, $r'=|\mathbf{r}'|$, and  $\vartheta$ as the angle between $\mathbf{r}$ and $\mathbf{r}'$. The coefficients $B_l$ are in general functions of $r'$. The first term above can  be expanded as \cite{Arfken} 
\begin{equation}
\label{eq:DHexpansion}
\frac{\exp(-\kappa|\mathbf{r}-\mathbf{r}'|)}{|\mathbf{r}-\mathbf{r}'|}=\kappa\sum_{l=0}^{\infty}(2l+1)i_l(\kappa r_<)k_l(\kappa r_>)P_l(\cos\vartheta),
\end{equation}
in which $i_l(\cdot)$ is  the modified spherical Bessel function of the first kind and $r_<$ and $r_>$ denote the smaller and larger values of $r$ and $r'$. Since the potential on the metallic sphere is constant and does not depend on $\vartheta$, and using  $r_<=r=R_0$ and $r_>=r'$, we find
\begin{equation}
\begin{array}{lr}
C=B_0k_0(\kappa R_0)+\frac{\kappa }{4\pi\varepsilon\varepsilon_0}i_0(\kappa R_0)k_0(\kappa r'), & \textrm{ for } l=0,\\
\\
0=B_lk_l(\kappa R_0)+\frac{\kappa }{4\pi\varepsilon\varepsilon_0}(2l+1)i_l(\kappa R_0)k_l(\kappa r'), &\textrm{ for } l>0.
\end{array}
\end{equation}
and, hence, 
\begin{equation}
\begin{array}{lr}
B_0=\frac{C}{k_0(\kappa R_0)}-\frac{\kappa }{4\pi\varepsilon\varepsilon_0}\frac{i_0(\kappa R_0)}{k_0(\kappa R_0)}k_0(\kappa r'),\\
\\
B_l=-\frac{\kappa }{4\pi\varepsilon\varepsilon_0}(2l+1)\frac{i_l(\kappa R_0)}{k_l(\kappa R_0)}k_l(\kappa r'), 
\end{array}
\end{equation}
which give the solution in the outside region, $r,r'\geq R_0$, as
\begin{equation}
\label{eq:dhimage}
G(\mathbf{r},\mathbf{r}')=C\frac{k_0(\kappa r)}{k_0(\kappa R_0)}-\frac{\kappa}{4\pi\varepsilon\varepsilon_0}\sum_{l=0}^{\infty}(2l+1)\frac{i_l(\kappa R_0)}{k_l(\kappa R_0)}k_l(\kappa r')k_l(\kappa r)P_l(\cos\vartheta)+\frac{1}{4\pi\varepsilon\varepsilon_0}\frac{\exp(-\kappa|\mathbf{r}-\mathbf{r}'|)}{|\mathbf{r}-\mathbf{r}'|}.
\end{equation}
The  constant $C$ can be fixed by using the fact that the metallic NP is assumed to be electroneutral; hence, using Gauss's law and after straightforward manipulations, we find
\begin{eqnarray}
C = \frac{\kappa}{4\pi\varepsilon\varepsilon_0}\frac {k_0(\kappa r')}{k_0'(\kappa R_0)}
\left\{ i_0(\kappa R_0)k_0'(\kappa R_0)-i_0'(\kappa R_0)k_0(\kappa R_0) \right\} = \frac{1}{4\pi\varepsilon\varepsilon_0}\frac{\rme^{-\kappa(r'-R_0)}}{ r'(1+\kappa R_0)}, 
\end{eqnarray}
with the explicit expressions 
\begin{equation}
i_0(x)=\frac{\sinh x}{x},\quad
i_0'(x)=\frac{x\cosh x-\sinh x}{x^2},\quad
k_0(x)=\frac{\rme^{-x}}{x},\quad
k_0'(x)=-\frac{(1+x)\rme^{-x}}{x^2}.
\end{equation}
This gives the final expression for the Green's function as 
\begin{equation}
G(\mathbf{r},\mathbf{r}')=G_0(\mathbf{r},\mathbf{r}')+G_{im}(\mathbf{r},\mathbf{r}'), 
\end{equation}
where $G_{im}(\mathbf{r},\mathbf{r}')$ is the contribution representing salt/dielectric image effects,  
\begin{equation}
G_{im}(\mathbf{r},\mathbf{r}')=-\frac{\kappa}{4\pi\varepsilon\varepsilon_0}\sum_{l=0}^\infty(2l+1)\frac{i_l(\kappa R_0)}{k_l(\kappa R_0)}k_l(\kappa r)k_l(\kappa r')P_l(\cos\vartheta)+\frac{\kappa^2 R_0\,\rme^{2\kappa R_0}}{4\pi\varepsilon\varepsilon_0(1+\kappa R_0)}k_0(\kappa r)k_0(\kappa r').
\end{equation}

\section{Hamiltonian of the model VLP}

In the VLP model used in the main text, the  charge distribution of the inner and outer spherical shells (of radii $R_1$ and $R_2$) can formally be expressed as 
\begin{equation}
\rho_\sigma(\mathbf{r})=\sum_{\alpha=1}^2\sigma_\alpha(\mathbf{r})=\sum_{\alpha=1}^2\sigma_\alpha\delta(r-R_\alpha).
\end{equation}
Other explicit charges in the system include multivalent ions each of charge $q$ located at position $\mathbf{r}_i$, giving the local charge distribution function,
\begin{equation}
\rho(\mathbf{r})=\sum_{i=1}^Nq\delta(\mathbf{r}-\mathbf{r}_i).
\end{equation}
The Hamiltonian associated with electrostatic interactions in the system can in general be written as
\begin{equation}
H=\frac{1}{2}\sum_{i,j=1}^Nq^2G(\mathbf{r}_i,\mathbf{r}_j)+\sum_{i=1}^N q\int\mathrm{d}^3\mathbf{r}\rho_\sigma(\mathbf{r})G(\mathbf{r},\mathbf{r}_i)+\frac{1}{2}\iint\mathrm{d}^3\mathbf{r}\,\mathrm{d}^3\mathbf{r}'\,\rho_\sigma(\mathbf{r})G(\mathbf{r},\mathbf{r}')\rho_\sigma(\mathbf{r}').
\end{equation}

Let us first focus on the case of only one multivalent ion in the system positioned at $\mathbf{b}$ (note that multivalent ion positions are restricted to remain outside the inner shell, i.e., $b=|{\mathbf b}|>R_1$. We will thus have 
\begin{equation}\label{eq:H_Sents}
H=\frac{q^2}{2}G(\mathbf{b},\mathbf{b})+q\int\mathrm{d}^3\mathbf{r}\rho_\sigma(\mathbf{r})G(\mathbf{r},\mathbf{b})+
\frac{1}{2}\iint\mathrm{d}^3\mathbf{r}\,\mathrm{d}^3\mathbf{r}'\,\rho_\sigma(\mathbf{r})G(\mathbf{r},\mathbf{r}')\rho_\sigma(\mathbf{r}')\equiv H_{im}+H_{\sigma_\alpha}+H_{\sigma\sigma}.
\end{equation}
The first term in Eq. \eqref{eq:H_Sents} is the self-energy of the multivalent ion and its image interaction. We subtract the redundant (infinite) vacuum self-energy of the multivalent ion, and the ion-image interaction term is found as  
\begin{equation}
H_{im}=-\frac{q^2\kappa }{8\pi\varepsilon\varepsilon_0}\sum_{l=0}^\infty(2l+1)\frac{i_l(\kappa R_0)}{k_l(\kappa R_0)}k_l^2(\kappa b)+\frac{q^2\kappa^2 R_0\,\rme^{2\kappa R_0}}{8\pi\varepsilon\varepsilon_0(1+\kappa R_0)}k_0^2(\kappa b).
\end{equation}
The second term in Eq. \eqref{eq:H_Sents} is the interaction between the ion and the surface charge, including both the direct, screened Coulomb (or DH), interaction and the image interaction. For the $\alpha$-th shell, it yields
\begin{equation}
H_{\sigma_\alpha}=\sum_{\alpha=1}^2 q\sigma_\alpha\int\mathrm{d}^3\mathbf{r}\,\delta(r-R_\alpha)G(\mathbf{r},\mathbf{b})=\sum_{\alpha=1}^2 q\sigma_\alpha\int r^2\mathrm{d}r\,\mathrm{d}\Omega\,\delta(r-R_\alpha)[G_0(\mathbf{r},\mathbf{b})+G_{im}(\mathbf{r},\mathbf{b})]\equiv H_{\sigma_\alpha}^{dir}+H_{\sigma_\alpha}^{im}.
\end{equation}
The direct interaction is
\begin{eqnarray}
\nonumber H_{\sigma_\alpha}^{dir}&=&\sum_{\alpha=1}^2 q\sigma_\alpha\int r^2\mathrm{d}r\,\mathrm{d}\Omega\,\delta(r-R_\alpha)G_0(\mathbf{r},\mathbf{b})\\
&=&\sum_{\alpha=1}^2 \frac{q\sigma_\alpha}{4\pi\varepsilon\varepsilon_0}\int r^2\mathrm{d}r\,\mathrm{d}\Omega\,\delta(r-R_\alpha)\frac{\rme^{-\kappa|\mathbf{r}-\mathbf{b}|}}{|\mathbf{r}-\mathbf{b}|}=\sum_{\alpha=1}^2 \frac{q\sigma_\alpha R_\alpha^2}{4\pi\varepsilon\varepsilon_0}\int\mathrm{d}\Omega\,\frac{\rme^{-\kappa\sqrt{R_\alpha^2+b^2-2R_\alpha b\cos\vartheta}}}{\sqrt{R_\alpha^2+b^2-2R_\alpha b\cos\vartheta}},
\end{eqnarray}
where $\vartheta$ is the angle between $\mathbf{r}$ and $\mathbf{b}$. Explicitly, we have
\begin{equation}
H_{\sigma_\alpha}^{dir}=\sum_{\alpha=1}^2 \frac{q\sigma_\alpha R_\alpha^2}{2\varepsilon\varepsilon_0}\int_1^{-1}-\mathrm{d}u\frac{\rme^{-\kappa\sqrt{R_\alpha ^2+b^2-2R_\alpha bu}}}{\sqrt{R_\alpha^2+b^2-2R_\alpha bu}}=\sum_{\alpha=1}^2 \frac{q\sigma_\alpha R_\alpha}{2\kappa b\varepsilon\varepsilon_0}\left(\rme^{-\kappa|b-R_\alpha|}-\rme^{-\kappa|b+R_\alpha|}\right).
\end{equation}
The image interaction part, on the other hand, is found as 
\begin{eqnarray}
\nonumber H_{\sigma_\alpha}^{im}&=&\sum_{\alpha=1}^2 q\sigma_\alpha\int r^2\mathrm{d}r\, \mathrm{d}\Omega\,\,\delta(r-R_\alpha)G_{im}(\mathbf{r},\mathbf{b})\\
&=&\sum_{\alpha=1}^2 -\frac{\kappa q\sigma_\alpha R_\alpha^2}{4\pi\varepsilon\varepsilon_0}\int\mathrm{d}\Omega\, \left( \sum_{l=0}^\infty(2l+1)\frac{i_l(\kappa R_0)}{k_l(\kappa R_0)}k_l(\kappa R_\alpha)k_l(\kappa b)P_l(\cos\vartheta)-\frac{\kappa R_0\,\rme^{2\kappa R_0}}{1+\kappa R_0}k_0(\kappa R_\alpha)k_0(\kappa b)\right), 
\end{eqnarray}
or, 
\begin{equation}
H_{\sigma_\alpha}^{im}=\sum_{\alpha=1}^2 -\frac{\kappa q\sigma_\alpha R_\alpha^2}{2\varepsilon\varepsilon_0}\left(\sum_{l=0}^\infty(2l+1)\frac{i_l(\kappa R_0)}{k_l(\kappa R_0)}k_l(\kappa R_\alpha)k_l(\kappa b)\int_{1}^{-1}-\mathrm{d}uP_l(u)-\int_0^\pi\sin\vartheta d\vartheta \frac{\kappa R_0\,\rme^{2\kappa R_0}}{1+\kappa R_0}k_0(\kappa R_\alpha)k_0(\kappa b)\right).
\end{equation}
The integral over Legendre functions is non-zero only for $l=0$, leaving us with
\begin{eqnarray}
\nonumber H_{\sigma_\alpha}^{im}&=&\sum_{\alpha=1}^2 -\frac{\kappa q\sigma_\alpha R_\alpha^2}{\varepsilon\varepsilon_0}\left(\frac{i_0(\kappa R_0)}{k_0(\kappa R_0)}k_0(\kappa R_\alpha)k_0(\kappa b)-\frac{\kappa R_0\,\rme^{2\kappa R_0}}{1+\kappa R_0}k_0(\kappa R_\alpha)k_0(\kappa b)\right)\\
&=&\sum_{\alpha=1}^2 -\frac{\kappa q\sigma_\alpha R_\alpha^2}{\varepsilon\varepsilon_0}\left(\rme^{\kappa R_0}\sinh \kappa R_0\frac{\rme^{-\kappa R_\alpha}}{\kappa R_\alpha}\frac{\rme^{-\kappa b}}{\kappa b}-\frac{\kappa R_0\,\rme^{2\kappa R_0}}{1+\kappa R_0}\frac{\rme^{-\kappa R_\alpha}}{\kappa R_\alpha}\frac{\rme^{-\kappa b}}{\kappa b}\right).
\end{eqnarray}
The net contribution from the second term in Eq. (\ref{eq:H_Sents}) is thus obtained as 
\begin{equation}
H_{\sigma_\alpha}=\sum_{\alpha=1}^2\left[\frac{q\sigma_\alpha R_\alpha}{2\kappa b\varepsilon\varepsilon_0}\left(\rme^{-\kappa|b-R_\alpha|}-\rme^{-\kappa(b+R_\alpha)}\right)+\frac{q\sigma_\alpha R_\alpha}{2\kappa b\varepsilon\varepsilon_0}\rme^{-\kappa(b+R_\alpha)}\left(1+\rme^{2\kappa R_0}\frac{\kappa R_0-1}{\kappa R_0+1}\right)\right].
\end{equation}
For the last part of Eq. \eqref{eq:H_Sents}, which gives the contribution from surface-surface interaction (including the relevant image effects), we can write
\begin{equation}
H_{\sigma\sigma}=\sum_{\alpha,\beta=1}^2 \frac{1}{2}\iint\mathrm{d}^3\mathbf{r}\,\mathrm{d}^3\mathbf{r}'\sigma_\alpha\delta(r-R_\alpha)[G_0(\mathbf{r},\mathbf{r}')+G_{im}(\mathbf{r},\mathbf{r}')]\sigma_\beta\delta(r'-R_\beta)\equiv H_{\sigma\sigma}^{dir}+H_{\sigma\sigma}^{im}.
\end{equation}
The direct interaction part here is give by
\begin{equation}
H_{\sigma\sigma}^{dir}=\sum_{\alpha,\beta=1}^2\frac{1}{2}\iint\mathrm{d}^3\mathbf{r}\,\mathrm{d}^3\mathbf{r}'\sigma_\alpha\delta(r-R_\alpha)G_0(\mathbf{r},\mathbf{r}')\sigma_\beta\delta(r'-R_\beta),
\end{equation}
or, 
\begin{equation}
H_{\sigma\sigma}^{dir}=\sum_{\alpha,\beta=1}^2\frac{\sigma_\alpha\sigma_\beta}{2}\iint r^2\mathrm{d}r\,\mathrm{d}\Omega\, r'^2\mathrm{d}r' \mathrm{d}\Omega\,'\delta(r-R_\alpha)\delta(r'-R_\beta)\frac{\rme^{-\kappa\dist}}{4\pi\varepsilon\varepsilon_0\dist}, 
\end{equation}
giving 
\begin{equation}
H_{\sigma\sigma}^{dir}=\sum_{\alpha,\beta=1}^2\frac{\sigma_\alpha\sigma_\beta R_\alpha^2R_\beta^2}{8\pi\varepsilon\varepsilon_0}\int\mathrm{d}\Omega\,\int\mathrm{d}\Omega\,'\frac{\rme^{-\kappa\sqrt{R_\alpha^2+R_\beta^2-2R_\alpha R_\beta\cos\vartheta}}}{\sqrt{R_\alpha^2+R_\beta^2-2R_\alpha R_\beta\cos\vartheta}}.
\end{equation}
The first angular integration above can be done straightforwardly, and since the result is independent of the angle between the two vectors, the second integration only yields a constant. Thus,
\begin{equation}
H_{\sigma\sigma}^{dir}=\sum_{\alpha,\beta=1}^2\frac{\sigma_\alpha\sigma_\beta R_\alpha^2R_\beta^2}{8\pi\varepsilon\varepsilon_0}4\pi\frac{2\pi}{\kappa R_\alpha R_\beta}\left(\rme^{-\kappa|R_\alpha-R_\beta|}-\rme^{-\kappa(R_\alpha+R_\beta)}\right)=\sum_{\alpha,\beta=1}^2\frac{\pi\sigma_\alpha\sigma_\beta R_\alpha R_\beta}{\kappa\varepsilon\varepsilon_0}\left(\rme^{-\kappa|R_\alpha-R_\beta|}-\rme^{-\kappa(R_\alpha+R_\beta)}\right).
\end{equation}
The image interaction part, on the other hand, is obtained as 
\begin{equation}
H_{\sigma\sigma}^{im}=\sum_{\alpha,\beta=1}^2\frac{1}{2}\iint\mathrm{d}^3\mathbf{r}\,\mathrm{d}^3\mathbf{r}'\sigma_\alpha\delta(r-R_\alpha)G_{im}(\mathbf{r},\mathbf{r}')\sigma_\beta\delta(r'-R_\beta), 
\end{equation}
or, similarly as before, 
\begin{eqnarray}
\nonumber H_{\sigma\sigma}^{im}&=&\sum_{\alpha,\beta=1}^2-\frac{\kappa\sigma_\alpha\sigma_\beta R_\alpha^2R_\beta^2}{8\pi\varepsilon\varepsilon_0}\\
\nonumber &&\times\left(\;\sum_l(2l+1)\frac{i_l(\kappa R_0)}{k_l(\kappa R_0)}k_l(\kappa R_\alpha)k_l(\kappa R_\beta)\int\mathrm{d}\Omega\,'\int\mathrm{d}\Omega\, P_l(\cos\vartheta)-\int\mathrm{d}\Omega\,'\int\mathrm{d}\Omega\,\frac{\kappa R_0\,\rme^{2\kappa R_0}}{1+\kappa R_0}k_0(\kappa R_\alpha)k_0(\kappa R_\beta)\;\right)\\
\nonumber &=&\sum_{\alpha,\beta=1}^2-\frac{\kappa\sigma_\alpha\sigma_\beta R_\alpha^2R_\beta^2}{8\pi\varepsilon\varepsilon_0}\\
&&\times\left(\sum_l(2l+1)\frac{i_l(\kappa R_0)}{k_l(\kappa R_0)}k_l(\kappa R_\alpha)k_l(\kappa R_\beta)4\pi(4\pi\delta_{l0})-(4\pi)(4\pi)\frac{\kappa R_0\,\rme^{2\kappa R_0}}{1+\kappa R_0}k_0(\kappa R_\alpha)k_0(\kappa R_\beta)\right),
\end{eqnarray}
from which we obtain
\begin{eqnarray}
\nonumber H_{\sigma\sigma}^{im}&=&\sum_{\alpha,\beta=1}^2-\frac{2\pi\kappa\sigma_\alpha\sigma_\beta R_\alpha^2R_\beta^2}{\varepsilon\varepsilon_0}\left(\frac{i_0(\kappa R_0)}{k_0(\kappa R_0)}k_0(\kappa R_\alpha)k_0(\kappa R_\beta)-\frac{\kappa R_0\,\rme^{2\kappa R_0}}{1+\kappa R_0}k_0(\kappa R_\alpha)k_0(\kappa R_\beta)\right)\\
&=&\sum_{\alpha,\beta=1}^2\frac{\pi\sigma_\alpha\sigma_\beta R_\alpha R_\beta}{\kappa\varepsilon\varepsilon_0}\left(1+\rme^{2\kappa R_0}\frac{\kappa R_0-1}{\kappa R_0+1}\right) \rme^{-\kappa(R_\alpha+R_\beta)}. 
\end{eqnarray}
Hence, 
\begin{equation}
H_{\sigma\sigma}=\sum_{\alpha,\beta=1}^2\frac{\pi\sigma_\alpha\sigma_\beta R_\alpha R_\beta}{\kappa\varepsilon\varepsilon_0}\left(\rme^{-\kappa|R_\alpha-R_\beta|}-\rme^{-\kappa(R_\alpha+R_\beta)}\right)+\sum_{\alpha,\beta=1}^2\frac{\pi\sigma_\alpha\sigma_\beta R_\alpha R_\beta}{\kappa\varepsilon\varepsilon_0}\left(1+\rme^{2\kappa R_0}\frac{\kappa R_0-1}{\kappa R_0+1}\right) \rme^{-\kappa(R_\alpha+R_\beta)}.
\end{equation}
Putting the three terms contributing to the Hamiltonian together, $H=H_{im}+H_{\sigma_\alpha}+H_{\sigma\sigma}$, we have 
\begin{eqnarray}
\nonumber H&=&-\frac{\kappa q^2}{8\pi\varepsilon\varepsilon_0}\sum_{l=0}^\infty(2l+1)\frac{i_l(\kappa R_0)}{k_l(\kappa R_0)}k_l^2(\kappa b)+\frac{\kappa^2 R_0\,\rme^{2\kappa R_0} q^2}{8\pi\varepsilon\varepsilon_0(1+\kappa R_0)}k_0^2(\kappa b)\\
\nonumber&&+\frac{q}{2\kappa b\varepsilon\varepsilon_0}\sum_{\alpha=1}^2\sigma_\alpha R_\alpha\left[\left(\rme^{-\kappa|b-R_\alpha|}-\rme^{-\kappa(b+R_\alpha)}\right)+ \rme^{-\kappa(b+R_\alpha)}\left(1+\rme^{2\kappa R_0}\frac{\kappa R_0-1}{\kappa R_0+1}\right)\right]\\
&&+\frac{\pi}{\kappa\varepsilon\varepsilon_0}\sum_{\alpha,\beta=1}^2\sigma_\alpha\sigma_\beta R_\alpha R_\beta \left[\left(\rme^{-\kappa|R_\alpha-R_\beta|}-\rme^{-\kappa(R_\alpha+R_\beta)}\right)+ \rme^{-\kappa(R_\alpha+R_\beta)}\left(1+\rme^{2\kappa R_0}\frac{\kappa R_0-1}{\kappa R_0+1}\right)\right].
\end{eqnarray}
%

When we have $N$ multivalent ions, the Hamiltonian can straightforwardly be expressed as 
\begin{eqnarray}\label{eq:Hamiltonian}
\nonumber H&=&\sum_{i=1}^N-\frac{\kappa q^2}{8\pi\varepsilon\varepsilon_0}\sum_{l=0}^\infty(2l+1)\frac{i_l(\kappa R_0)}{k_l(\kappa R_0)}k_l^2(\kappa r_i)+\sum_{i=1}^N\frac{\kappa^2 R_0\,\rme^{2\kappa R_0} q^2}{8\pi\varepsilon\varepsilon_0(1+\kappa R_0)}k_0^2(\kappa r_i)\\
\nonumber&+&\sum_{i=1}^N\frac{q}{2\kappa r_i\varepsilon\varepsilon_0}\sum_{\alpha=1}^2\sigma_\alpha R_\alpha\left[\left(\rme^{-\kappa|r_i-R_\alpha|}-\rme^{-\kappa(r_i+R_\alpha)}\right)+ \rme^{-\kappa(r_i+R_\alpha)}\left(1+\rme^{2\kappa R_0}\frac{\kappa R_0-1}{\kappa R_0+1}\right)\right]\\
\nonumber&+&\frac{\pi}{\kappa\varepsilon\varepsilon_0}\sum_{\alpha,\beta=1}^2\sigma_\alpha\sigma_\beta R_\alpha R_\beta\left[\left(\rme^{-\kappa|R_\alpha-R_\beta|}-\rme^{-\kappa(R_\alpha+R_\beta)}\right)+ \rme^{-\kappa(R_\alpha+R_\beta)}\left(1+\rme^{2\kappa R_0}\frac{\kappa R_0-1}{\kappa R_0+1}\right)\right]\\
\nonumber &+&\sum_{i>j=1}^N\frac{q^2}{4\pi\varepsilon\varepsilon_0}\frac{\rme^{-\kappa|\mathbf{r}_i-\mathbf{r}_j|}}{|\mathbf{r}_i-\mathbf{r}_j|}\\
&-&\sum_{i>j=1}^N\frac{\kappa q^2}{4\pi\varepsilon\varepsilon_0}\sum_{l=0}^\infty(2l+1)\frac{i_l(\kappa R_0)}{k_l(\kappa R_0)}k_l(\kappa r_i)k_l(\kappa r_j)P_l(\cos\vartheta)
+\sum_{i>j=1}^N\frac{\kappa^2 R_0 \,\rme^{2\kappa R_0}q^2}{4\pi\varepsilon\varepsilon_0(1+\kappa R_0)}k_0(\kappa r_i)k_0(\kappa r_j).
\end{eqnarray}
This completes the derivation of the expressions given in Eqs. (2)-(6) of the main text. 

\section{Osmotic (electrostatic) pressure on the outer shell}

In absence of a metallic core within the VLP,  the net electrostatic potential of the charged shells at $R_1$ and $R_2$ is obtained as
\begin{align}
\varphi_{1}(0\leq r\leq R_1)= \frac{\rme^{-\kappa R_1} R_1 \sigma_1+\rme^{-\kappa R_2} R_2 \sigma_2}{\varepsilon\varepsilon_0}\frac{\sinh \kappa r}{\kappa r},
\end{align}
\begin{align}
\varphi_{2}(R_1< r\leq R_2)= \frac{\rme^{-\kappa R_2}R_2\sigma_2}{\varepsilon\varepsilon_0}\frac{\sinh \kappa r}{\kappa r}+\frac{R_1\sigma_1 \sinh \kappa R_1}{\varepsilon\varepsilon_0}\frac{\rme^{-\kappa r}}{\kappa r},
\end{align}
\begin{align}
\varphi_{3}(r> R_2)=\frac{R_1\sigma_1 \sinh \kappa R_1+R_2\sigma_2 \sinh \kappa R_2}{\varepsilon\varepsilon_0}\frac{\rme^{-\kappa r}}{\kappa r}.
\end{align}
The free energy of the system in the absence of multivalent ions then follows standardly as 
\begin{align}
F_{DH}=&\;\frac{\pi}{\kappa \varepsilon\varepsilon_0} \left\{ \sigma_1^2 R_1^2(1-\rme^{-2\kappa R_1}) + \sigma_2^2 R_2^2(1-\rme^{-2\kappa R_2})+2\sigma_1\sigma_2 R_1 R_2(\rme^{-\kappa (R_2-R_1)}-\rme^{-\kappa(R_1+R_2)})\right\}.
\end{align}
and the osmotic pressure as 
\begin{align}\label{eq:PDH_No}
P_{DH}=\;\frac{\sigma_2^2}{2\varepsilon \varepsilon_0} \left\{ \frac{1}{\kappa R_2}-\rme^{-2\kappa R_2}\left(1+\frac{1}{\kappa R_2}\right) \right\}+\frac{\sigma_2 \sigma_1}{2\varepsilon \varepsilon_0} \frac{R_1}{R_2}\left\{ \rme^{-\kappa (R_2-R_1)}-\rme^{-\kappa (R_1+R_2)}\right\}\left(1+\frac{1}{\kappa R_2}\right).
\end{align}

The contribution of multivalent ions to the osmotic pressure follows as (see Refs. \cite{leili1b,leili2b})
\begin{equation}
P_q=-\left\langle\sum_{i=1}^N q\frac{\partial \varphi(\mathbf{r}_i)}{\partial V_2}\bigg|_{Q_2}\right\rangle,
\end{equation}
where $V_2=4\pi R_2^3/3$ is the volume of the outer shell and the partial derivative is taken at fixed value of the total surface charge of this shell, i.e.,  ${Q_2}=4\pi R_2^2 \sigma_2$. This contribution can directly be calculated by noting that 
\begin{equation}
\frac{\partial\varphi_2(R_1< r\leq R_2)}{\partial V_2}\bigg|_{Q_{2}}=\;\frac{\sigma_2 \sinh \kappa r}{4\pi\varepsilon\varepsilon_0\kappa r R_2^2}(1+\kappa R_2)\rme^{-\kappa R_2},
\end{equation}
and
\begin{equation}
\frac{\partial\varphi_3(r> R_2)}{\partial V_2}\bigg|_{Q_{2}}=\;\frac{\sigma_2 \rme^{-\kappa r}}{4\pi\varepsilon\varepsilon_0\kappa r R_2^2}(\kappa R_2 \cosh \kappa R_2 - \sinh \kappa R_2). 
\end{equation}

In presence of a metallic core, the potential derivative is found as 
\begin{align}\label{eq:d_phi_V2}
\frac{\partial\varphi(\mathbf{r}_i)}{\partial V_2}\bigg|_{Q_{2}}=\;\sum_{i=1}^N \frac{\sigma_2}{8\pi\varepsilon \varepsilon_0\kappa R_2^2}
\left[\frac{\rme^{-\kappa|r_i-R_2|}}{r_i}\left(\kappa R_2\,{\mathrm{sgn}}(r_i-R_2)-1\right)-\frac{\rme^{-\kappa (r_i+R_2-2R_0)}}{r_i}(1+\kappa R_2)\left(\frac{\kappa R_0-1}{\kappa R_0+1}\right)\right],
\end{align}
which follows from the second term in Eq. \eqref{eq:Hamiltonian}. This expression can be used to construct the contribution of multivalent ions to the osmotic pressure in this case, reproducing Eqs. (12) and (13) in the main text. Also, the third term in Eq. \eqref{eq:Hamiltonian}, can be used to obtain Eqs. (9)-(11) in the main text, giving 
\begin{align}\label{eq:PDH_MN}
\nonumber P_{DH}=&\;\frac{\sigma_2^2}{2\varepsilon \varepsilon_0} \left\{ \frac{1}{\kappa R_2}+\rme^{-2\kappa (R_2-R_0)}\left(\frac{\kappa R_0-1}{\kappa R_0+1}\right)\left(1+\frac{1}{\kappa R_2}\right) \right\}\\
+&\frac{\sigma_2 \sigma_1}{2\varepsilon \varepsilon_0} \frac{R_1}{R_2}\left\{ \rme^{-\kappa (R_2-R_1)}+\rme^{-\kappa (R_1+R_2-2R_0)}\left(\frac{\kappa R_0-1}{\kappa R_0+1}\right)\right\}\left(1+\frac{1}{\kappa R_2}\right) 
.
\end{align}

\end{widetext}

\end{document}